\documentclass[a4paper,11pt]{article}
\usepackage{jheppub} 
\usepackage{lineno}
\usepackage{upgreek}
\usepackage{float, extarrows, tikz-cd}
\usepackage{graphicx}
\usepackage{slashed}
\usepackage{tabularx,ragged2e}
\usepackage{amssymb}
\usepackage{subcaption}
\usepackage{amsmath,amssymb}
\usepackage{slashed}
\usepackage{caption}
\usepackage{xcolor}
\usepackage{dsfont}
\usepackage{verbatim}
\usepackage{mathtools, xcolor,ytableau, amsfonts,tikz}
\usepackage{graphicx}
\usepackage{physics}
\usepackage{simpler-wick}
\usepackage{microtype}
\usepackage{graphicx}
\usepackage[nobottomtitles*]{titlesec}
\newcolumntype{C}{>{\Centering\arraybackslash}X}

\usepackage{placeins}

\usepackage{xcolor}
\definecolor{amethyst}{rgb}{0.54, 0.17, 0.89}
\definecolor{coral}{rgb}{1.0, 0.3, 0.4}

\numberwithin{equation}{section}

\newcommand\ov{\over}
\newcommand\p{\ensuremath{\partial}}
\newcommand{\es}[2] {\begin{equation} \label{#1} \begin{split} #2 \end{split} \end{equation}}

\def\<{\langle}
\def\>{\rangle}

\def\mop#1{\mathop{\rm #1}\nolimits}
\def\vol{\mop{vol}}
\def\AreaLaw{{\rm Area}(\p A)}

\title{\boldmath Time Evolution of Multi-Party Entanglement Signals}
\author[a,b,c]{Vijay Balasubramanian}        
\author[c]{Hanzhi Jiang,}        
\author[d]{and Simon F. Ross}    

                \affiliation[a]{Department of Physics and Astronomy, University of Pennsylvania, Philadelphia, PA 19104, U.S.A.} 

                \affiliation[b]{Theoretische Natuurkunde, Vrije Universiteit Brussel and International Solvay Institutes, Pleinlaan 2, B-1050 Brussels, Belgium} 

                \affiliation[c]{Rudolf Peierls Centre for Theoretical Physics, University of Oxford, Oxford OX1 3PU, U.K.} 

                \affiliation[d]{Centre for Particle Theory, Department of Mathematical Sciences, Durham University, Durham DH1 3LE, U.K.} 

\emailAdd{vijay@physics.upenn.edu}
\emailAdd{hanzhi.jiang@physics.ox.ac.uk}     
\emailAdd{s.f.ross@durham.ac.uk}   

\abstract{
We study the real-time dynamics of multi-party entanglement signals in chaotic quantum many-body systems including but not necessarily restricted to holographic conformal field theories. We find that scrambling dynamics generates multiparty entanglement with rich structure including: (a) qualitatively different dynamical behaviours for different signals, likely reflecting different dynamics for different kinds of entanglement patterns, (b) discontinuities possibly indicating dynamical phase transitions in the entanglement structure, (c) transient and non-monotonic multiparty entanglement signals,  and (d) periods during which the extensive entanglement of some regions is entirely multipart\textbf{}ite.  Our main technical tool is the membrane theory of entanglement dynamics.

}

\begin{document}
\maketitle
\flushbottom

\section{Introduction}
\label{sec:intro}
Entanglement is a fundamental concept of quantum systems, underpinning phenomena ranging from thermalization of quantum many-body systems~\cite{Deutsch:1991msp,Srednicki:1994mfb,Rigol:2007juv} to the emergence of spacetime~\cite{VanRaamsdonk:2010pw,Maldacena:2013xja} in holography. Most work focuses on bipartite entanglement, where we divide a system into two parts, and study quantities like the von Neumann entropy of one part.

Systems with many degrees of freedom can also display intricate multiparty entanglement patterns. Multiparty entanglement has rich structure, and we do not have a good general description of the forms it can take (see the review \cite{Walter:2016lgl}). Some information about multiparty entanglement can be obtained by considering entanglement across different bipartitions of a system; for example the triple information constructed from such data provides a signal of four-party entanglement \cite{Cui:2018dyq}. Further insight can be obtained from the {\it reflected entropy} calculated from canonical purification of appropriately chosen subsystems \cite{Dutta:2019gen}. This quantity is thought to be computed holographically in terms of an {\it Entanglement Wedge Cross-Section} (EWCS) \cite{Dutta:2019gen}. See \cite{Bao:2025nbz} for a derivation in the context of the AdS$_3$/CFT$_2$ duality. As we review below, the difference between reflected entropy and mutual information provides a signal of three-party entanglement \cite{Akers:2019gcv,Hayden:2021gno}, as does the {\it genuine multi-entropy}  \cite{Gadde:2023zzj,Harper:2024ker,Iizuka:2025ioc,Iizuka:2025pqq}. Another signal of tripartite entanglement was recently proposed in \cite{Bao:2025psl}. By ``signal'' we mean here a quantity whose non-vanishing confirms the presence of multi-party entanglement.  The converse need not be true.  If one of the signals we consider vanishes, there may still be some multi-party entanglement; it is just not detected by the signal in question.  

Recently, the authors of \cite{Balasubramanian:2024ysu} initiated a study of multiparty entanglement signals for general numbers of parties.  They used these signals to show that states with holographically dual geometries contain significant amounts of multiparty entanglement.  Using these methods \cite{Balasubramanian:2025hxg} derived an inequality showing that 
purely GHZ-like entanglement between any number of parties is forbidden in holography,\footnote{This inequality was  shown~\cite{Akella:2025owv} to be either violated or saturated by stabiliser states~\cite{Gottesman:1997zz}.}  consistent with arguments for three-party GHZ-like states in \cite{Hayden:2021gno}. The discussion in these works, and in the preceding analysis of  \cite{Balasubramanian:2014hda},  was restricted to static or time-symmetric cases, such as vacuum AdS$_3$ and multiboundary wormholes.  Here we consider the time evolution of multiparty entanglement signals.


We will consider time evolution in chaotic systems in a late time, large region limit, in which entanglement dynamics is conjectured to be universally captured by an effective description in terms of a certain minimal membrane~\cite{Nahum:2016muy,Jonay:2018yei,Mezei:2018jco}.\footnote{For the special case of two dimensional holographic CFT, the infinite dimensional conformal symmetry requires a slight generalization of the membrane theory to include an additional degree of freedom~\cite{Jiang:2024tdj}.} This scenario includes, but is not limited to, simple time-dependent holographic states dual to asymptotically AdS black holes~\cite{Balasubramanian:2011ur,Shenker:2013pqa,Roberts:2014ifa}. Likewise, the membrane picture has been used to show that the holographic entropy cone inequalities \cite{Bao:2015bfa} that apply to the AdS vacuum also apply to the late time, large region limit of thermal holographic states \cite{Bao:2018wwd}.  Overall, the simplicity and generality of the membrane description allow us to extract information about the time-dependent behaviour of multiparty entanglement signals.

The evolution of the triple information in time-dependent holographic states dual to asymptotically AdS black holes was previously considered in \cite{Balasubramanian:2011at}.\footnote{In the precise setup of~\cite{Balasubramanian:2011at} the membrane theory does not apply because they do not examine a late time, large region limit.}  For the analysis there it was sufficient to use the HRT formula \cite{Hubeny:2007xt} for entanglement entropy of intervals in holographic theories because the triple information is constructed from such quantities.  To construct the entanglement signals studied in \cite{Balasubramanian:2024ysu} we need to additionally compute reflected entropies \cite{Dutta:2019gen,Akers:2019gcv,Hayden:2021gno,Gadde:2023zzj,Harper:2024ker,Iizuka:2025ioc}.  The time evolution of reflected entropy was first studied in 2d CFT in~\cite{Kudler-Flam:2020url}. More recently, the membrane theory framework was enriched to enable computation of reflected entropy~\cite{Jiang:2024tdj}. We will use this method to study the dynamics of the multiparty signals considered in~\cite{Balasubramanian:2024ysu}.

Six sections follow. In Sec.~\ref{sec:MPEReview}, we review the multipartite entanglement signals discussed in~\cite{Balasubramanian:2024ysu}.  In Sec.~\ref{sec:Membrane} we review the membrane theory of entanglement dynamics~\cite{Mezei:2018jco,Jiang:2024tdj}. We describe the dynamics of three-party and four-party entanglement signals  in Secs.~\ref{sec:r3} and  \ref{sec:I3}. In Sec.~\ref{sec:higher}, we consider extensions to more parties. In Sec.~\ref{app:gen} we discuss multiparty entanglement signals in a generalised membrane theory describing two-dimensional conformal field theories.   We conclude with a discussion in  Sec.~\ref{sec:discussion}.

Overall, we find that scrambling dynamics generally produces  multiparty entanglement with rich structure that varies over time. Notably:
\begin{itemize}
\item {\it Different entanglement signals show different dynamical behaviours:} the fact that there are many distinct multiparty entanglement signals is useful, and perhaps necessary, because they likely capture different {\it kinds} of entanglement patterns.\footnote{See \cite{Verstraete:2002gqj} for a classification of four qubit entanglement patterns.} Indeed, we see in our examples 
that the different signals we consider have qualitatively different dynamical behaviour. We interpret them as capturing different aspects of the dynamics of multiparty entanglement. 
\item {\it Discontinuities:} signals based on bipartitions, such as the triple information, are always continuous, but signals based on the reflected entropy can have discontinuities. We will see that there can be dynamical discontinuities in the three-party entanglement in the membrane description. This suggests phase transitions in the entanglement structure of the state.
\item {\it Non-monotonicity:} we consider entanglement between large regions, while the initial state has entanglement at small scales. So our multiparty entanglement signals always vanish in the initial state.\footnote{When the state is dual to Vaidya-BTZ spacetime \cite{Balasubramanian:2010ce}, the results in~\cite{Balasubramanian:2024ysu} indicate that there are non-zero initial multiparty entanglement signals. But these initial signals are negligible compared to entropy in the late time and large region limit~\cite{Mezei:2018jco}.} Surprisingly, the subsequent evolution is not monotonic: after an initial increase, in some cases the entanglement signal decreases again.  In general, non-monotonicity in the multiparty entanglement signals does not necessarily imply non-monotonicity in the entanglement structures, as the signals only capture partial information of multipartite entanglement. Nevertheless, in Section~\ref{sec:r3twor} we will see a case where we have some evidence that the decrease in the signal is associated with a decrease in multiparty entanglement.

\item {\it Saturation of bounds:} the authors of  \cite{Balasubramanian:2024ysu} constructed bounds on  multiparty entanglement. We will see that in some cases, there are periods where the entanglement captured by the membrane description saturates one of these bounds, indicating that the extensive part of the entanglement of one of the regions is {\it entirely} multipartite. 
\end{itemize}

\section{Review of multi-party entanglement signals}
\label{sec:MPEReview}

In this section, we review some multi-party entanglement signals whose time evolution we will study. We will focus primarily on signals of three- and four-party entanglement, but we also include a brief discussion of higher-party entanglement signals. We focus on the signals discussed in~\cite{Balasubramanian:2024ysu}, but it would be interesting in the future to extend this to study the time-dependence of other signals, such as the genuine multi-entropy of \cite{Gadde:2023zzj,Harper:2024ker,Iizuka:2025ioc,Iizuka:2025pqq} and the new quantities defined in \cite{Bao:2025psl}.

We consider  multi-party entanglement in a pure state where the system is divided into $n$ disjoint subsystems. For $n=2$, we have a state $\ket{\psi}_{AB}$, and bipartite entanglement is measured by the von Neumann entropy $S(A)$ of the reduced density matrix $\rho_A = \mbox{tr}_B(\ket{\psi}_{AB} \bra{\psi}_{AB})$. We can also think of this in terms of the mutual information 
\begin{align}
    I(A:B)=S(A)+S(B)-S(AB),\label{MI}
\end{align}
as for a pure state $\ket{\psi}_{AB}$ we have $S(AB)=0$ and $S(A)=S(B)$, so $I(A:B)=2S(A)$.  

For three parties, we can trace the state $\ket{\psi}_{ABC}$ over one party, say $C$, to obtain the reduced density matrix $\rho_{AB}$. The mutual information \eqref{MI} for this reduced density matrix is sensitive to correlation between $A$ and $B$, which includes contributions from both bipartite entanglement between $A$ and $B$ and tripartite entanglement involving $A$, $B$ and $C$. To isolate the tripartite entanglement we need to combine this with another source of information on correlation between $A$ and $B$. A useful quantity is the reflected entropy
\begin{equation}\label{eq:RefEnt}
S_{R}(A:B) = S(\rho_{AA^*})\,, \quad \rho_{AA^*} = \Tr_{BB^*}\ket{\rho^{1/2}_{AB}}\bra{\rho^{1/2}_{AB}}\,,
\end{equation}
where $\ket{\rho^{1/2}_{AB}}\in \mathcal{H}_{ABA^*B^*}$ is the canonical purification of  $\rho_{AB}$~\cite{Dutta:2019gen}, and $A^*B^*$ are copies of $AB$. The doubled Hilbert space $\mathcal{H}_{ABA^*B^*}$ is constructed from mirror copies of the original system. This reflected entropy is also sensitive to both bipartite entanglement between $A$ and $B$ and tripartite entanglement involving $A$, $B$, and $C$. In the residual information
\begin{align}
    R_3(A:B)=S_R(A:B)-I(A:B),\label{R3}
\end{align}
the contribution from bipartite entanglement cancels, so the residual information is a signal of three-party entanglement: if $R_3 \neq 0$, the state $\ket{\psi}_{ABC}$ has tripartite entanglement between $A$, $B$, and $C$ \cite{Akers:2019gcv}. The residual information is provably non-negative \cite{Dutta:2019gen}, but it can nevertheless vanish on states with tripartite entanglement. For example, it vanishes for the GHZ state. 

Note that we chose to trace over $C$; tracing over $B$ or $A$ instead defines other signals of three-party entanglement, $R_3(A:C)$, $R_3(B:C)$. We will see later that these different signals can have qualitatively different behaviour, giving us independent information about aspects of the three-party entanglement. 

For a four party state $\ket{\psi}_{ABCD}$, we can obtain a signal of four-party entanglement by tracing over one of the parties, say $D$, and considering the triple information 
\begin{align}
    I_3(A:B:C)=S_A+S_B+S_C-S_{AB}-S_{AC}-S_{BC}+S_{ABC}\label{I3}
\end{align}
for the resulting reduced density matrix  \cite{Balasubramanian:2014hda}. This is permutation invariant among the four parties, which is made manifest by defining a four-party analogue $I_4(A:B:C:D)$ (see \eqref{ninf} for the general $n$-party definition), and noting that on pure states $I_4(A:B:C:D) = 2 I_3(A:B:C)$. This is analogous to the relation between the mutual information and the von Neumann entropy for two parties. The triple information is not sign-definite in general, although it is always negative on holographic states \cite{Hayden:2011ag}. 

In \cite{Balasubramanian:2024ysu}, another signal of four-party entanglement was defined using canonical purification. If we consider the reduced density matrix $\rho_{AB}$ obtained from $\ket{\psi}_{ABCD}$, using the canonical purification as above we can define the reflected entropy $S_R(A:B)$. Alternatively, if we reduce over $D$ we obtain a reduced density matrix $\rho_{ABC}$, with a canonical purification $\ket{\sqrt{\rho}}_{ABCA^*B^*C^*}$. Reducing this over $CC^*$ defines another reduced density matrix $\bar \rho_{AA^*BB^*}$, with a canonical purification $\ket{\sqrt{\bar{\rho}}}_{AA^*BB^*A_*A^*_*B_*B^*_*}$. Here, we use upper and lower $*$ to distinguish multiple copies of the spaces involved in the purifications.
Using the von Neumann entropy $S(AA^*A_*A^*_*)$ in this state, we can define the four-party residual entropy 
\begin{equation}
     Q_4=S(AA^*A_*A^*_*)-2S_R(A:B),  \label{Q4}
\end{equation}
which is non-zero only if the state has four-party entanglement. As with the residual information, this is not permutation invariant; we can define different signals by tracing out different subsystems. 

In \cite{Balasubramanian:2024ysu}, the residual information and triple information were generalised to higher numbers of parties. For even numbers of parties we have the $n$-information
\begin{equation} \label{ninf}
    I_n(A_1:\ldots:A_n) = - \sum_{i \leq n} (-1)^n S_i,  
\end{equation}
where $S_i$ is the permutation-invariant combination of all the entropies on sets of $i$ parties chosen from the $n$ parties.\footnote{Note that it has been recently shown that of these quantities, only the mutual information is monotonic until the partial trace \cite{Hernandez-Cuenca:2023iqh}. See also~\cite{Ju:2024hba,Ju:2025tgg} for further studies on $n$-information.} On pure states $I_n = 2 I_{n-1}$, where $I_{n-1}$ is evaluated on the reduced density matrix obtained by tracing out any one party. For $n$ odd $I_n=0$ on pure states, so the $n$-information is not helpful.  For odd $n$ we can define instead the $n$-residual information, by tracing out one party and taking the difference between $I_{n-1}$ evaluated on the canonical purification and $I_{n-1}$ on the density matrix, 
\begin{equation}
    R_n(A_1: \ldots : A_{n-1};A_n) = \frac{1}{2} I_{n-1}(A_1 A_1^*: \ldots :A_{n-1} A_{n-1}^*) - I_{n-1}(A_1: \ldots :A_{n-1}). 
\end{equation}
As with the residual information, this depends on which party we trace out, so there are $n$ independent signals. The $n$-information and the $n$-residual information for $n$ even and odd respectively are all signals of $n$-party entanglement; if they are non-zero the state has $n$-party entanglement. Beyond the cases discussed above nothing is known about their positivity properties.

\subsection{Multiparty entanglement signals in holography}
\label{sec:mphol}

If we consider states in a CFT with a holographic dual and take the subsystems to correspond to spatial subregions, the von Neumann entropy of any collection $X$ of subsystems is related to the area of an associated extremal surface $\gamma_X$ in the bulk by the RT~\cite{Ryu:2006bv,Ryu:2006ef}/HRT~\cite{Hubeny:2007xt} formula, 
\begin{equation}
 S(X) = \frac{A(\gamma_X)}{4G}.
\end{equation}
This expression can be used to calculate the $n$-information holographically. 

The reduced density matrix on $X$, $\rho_X$, is dual to the entanglement wedge (EW), which is the domain of development of a spacelike surface in the bulk whose boundary is the  region $X$ together with the HRT surface $\gamma_X$. The canonical purification of $\rho_X$ is then conjectured to be dual to a spacetime obtained from evolving a Cauchy surface obtained by gluing together two copies of this spacelike surface along $\gamma_X$ \cite{Dutta:2019gen,Engelhardt:2018kcs}. The von Neumann entropies in canonical purifications can then be obtained by applying the HRT formula to this doubled spacetime. This enables us to calculate the $n$-residual information and the four-party residual entropy holographically. In the case of the reflected entropy, the HRT surface in the doubled spacetime is twice the entanglement wedge cross-section, the extremal surface in the entanglement wedge which separates the $A$ and $B$ boundary subregions. 
  
\section{Calculating time-dependence using membranes}
\label{sec:Membrane}

\begin{figure}[htbp]
\centering
\includegraphics[width=.5\textwidth]{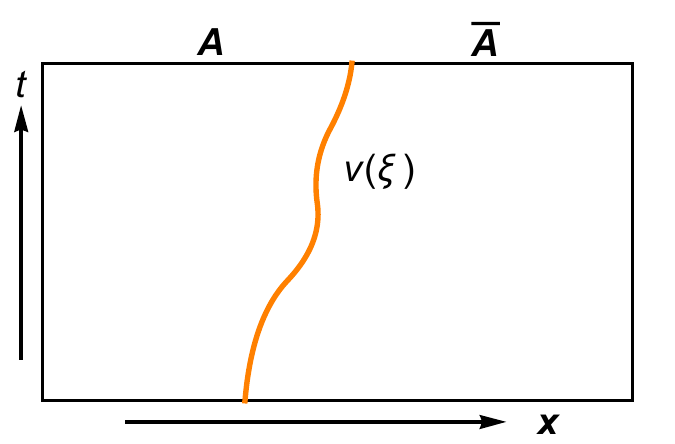}
\qquad
\caption{Cartoon of a minimal membrane (orange) extending in spacetime. This membrane computes  entanglement entropy between the subregion $A$ and its complement $\overline{A}$ at time $t$.
\label{fig:MembraneCartoon}}
\end{figure}

We are interested in studying the time-dependence of these multiparty entanglement signals. We will consider spatial subregions in an unbounded system. In the late time, large subregion limit, there is a general effective description of the dynamics  of entanglement and R\'enyi entropies in chaotic quantum many-body systems in terms of a minimal membrane~\cite{Nahum:2016muy,Jonay:2018yei,Mezei:2018jco}. In this effective description, the time-dependent entanglement entropy is computed by an action
\begin{equation}\label{MinMemb}
S=\min_{\vec{x}(\xi)} \int d^{d-1}\xi\, \sqrt{\abs{\gamma(\xi)}}\, \frac{s_\text{th}\, {\cal E}(\mathfrak{v}(\xi))}{\sqrt{1-\mathfrak{v}^2(\xi)}}\,.
\end{equation}
In~\eqref{MinMemb}, $\xi$ is the membrane worldvolume coordinate, $\vec{x}(\xi)$ is its embedding into spacetime, $\gamma$ is the induced metric on the membrane, $\mathfrak{v}(\xi)=\frac{d x(\xi)}{d t(\xi)}$ is the local transverse velocity of the membrane, and ${\cal E}(\mathfrak{v})$ is the velocity-dependent membrane tension function.
The minimisation is performed over all possible shapes of membranes that bound the entangling subregions at the  time $t=T$ at which we are computing  entanglement entropy. The quantity $s_\text{th}$ is a coarse-grained entropy density that is well-defined when local equilibrium has been reached, which in turn requires that $t, x \gg \beta$, where $\beta$ is the inverse temperature. In principle, $\gamma$ can be induced from any metric on which the theory is defined; in what follows, we will focus on the Minkowski metric. The membrane tension function ${\cal E}(\mathfrak{v})$ depends on the microscopic details of the specific theory under consideration, and has been derived for random quantum circuits~\cite{Nahum:2016muy,Jonay:2018yei,Zhou:2018myl}, Floquet circuits~\cite{Zhou:2019pob}, generalised dual-unitary circuits~\cite{Rampp:2023vah}, Brownian Hamiltonians~\cite{Vardhan:2024wxb}, and holographic conformal field theories~\cite{Mezei:2018jco}. In the holographic cases, ${\cal E}(\mathfrak{v})$ is specified by the geometry of the dual gravitational theories in the $t, x \gg \beta$ limit, as we will review shortly.  Finally, if there is initial bipartite entanglement~\cite{Jonay:2018yei} we must add an additional term in~\eqref{MinMemb} but  we will focus on cases with vanishing  initial (bipartite or multipartite) entanglement. See Fig.~\ref{fig:MembraneCartoon} for a sketch.



Suppose we consider a $d$-dimensional CFT, and  are interested in the late-time dynamics of  entanglement entropy for a region where one of the spatial coordinates, $x$ is restricted to a semi-infinite or finite region, in a state dual 
to an asymptotically AdS$_{d+1}$  black brane. We write the metric in infalling coordinates, 
\begin{equation}
ds^2 = \frac{1}{z^2}\left(-a(z)dv^2 -{2\ov b(z)}dvdz+dx^2+ d\vec{y}_{d-2}^2\right)\,,\label{blackBraneInf}
\end{equation}
where the conformal boundary is at $z=0$, and  AdS$_{d+1}$ asymptotics requires $a(0)=b(0)=1$. We assume there is a horizon at $z=1$, i.e., $a(1)=0$. A simple example is a neutral black brane, where $a(z)=1-z^d,\, b(z)=1$. We will either consider a semi-infinite region $x>0$, or a bounded one $x \in (0, \ell)$, for all $\vec y$.\footnote{We can also consider spherical or more general entangling subregions~\cite{Mezei:2018jco}.} 



First consider a semi-infinite region. From the dual gravity perspective, the leading dynamics of the entanglement entropy in the late time limit is determined by the portion of the HRT surface \cite{Hubeny:2007xt} that lies inside the horizon; for $d>2$, the HRT surfaces get stuck at a special extremal slice $z=z_*>1$ in the interior instead of moving deeper towards the singularity~\cite{Balasubramanian:2010ce,Balasubramanian:2011ur,Hartman:2013qma,Liu:2013qca,Liu:2013iza,Mezei:2016zxg,Mezei:2018jco,Jiang:2024tdj}.\footnote{ The $d=2$ case is different as discussed in  Sect.~\ref{app:gen}; nevertheless, the $d=2$ case with a relevant deformation becomes the same as the $d>2$ cases~\cite{Jiang:2024tdj}.}  This gives a linearly growing entanglement entropy, with a slope 
\begin{equation}
    v_E=\sqrt{-\frac{a(z)}{z^{2(d-1)}}}\Big|_{z=z_*}\,,\label{vE}
\end{equation}
called the entanglement velocity. This dynamics is reproduced by 
a vertical effective membrane (velocity $\mathfrak v(\xi)=0$) in the CFT with 
\begin{equation}
\label{AreaFunct2}
\begin{aligned}
S &= s_\text{th} \vol(\p A) \int dv \,  \mathcal{E}(0)\,, \qquad
\mathcal{E}(0) =\left. \sqrt{\frac{-a'(z)}{2(d-1)z^{2d-3}}}\right|_{z=z_*}\,, 
\end{aligned}
\end{equation}
where $\vol(\p A)$ is the volume in the transverse directions $\vec y$. In future expressions we will not write this (infinite) factor, so what we are actually describing is an entropy density along the $\vec y$ directions.  The bulk infalling time $v$ becomes the time coordinate on the membrane. As the infalling time agrees with the time coordinate on the boundary, this can be interpreted geometrically as identifying the membrane profile with the projection of the bulk HRT surface to the boundary along constant infalling time. Projecting more general HRT surfaces to the boundary in the same way will give an effective membrane picture with velocity $v(\xi) \in (0, v_B)$, where the butterfly velocity 
\begin{align}
    v_B=\sqrt{-\frac{a'(1)}{2(d-1)}}\,\label{vB}
\end{align}
is related to the out-of-time order correlator (OTOC)~\cite{Roberts:2014isa} and quantifies the speed at which chaos spreads in space. In general $v_E \leq v_B$~\cite{Mezei:2016zxg}.

The HRT surfaces have infinite area as $z\to 0$, reflecting the UV divergences of  entanglement entropy in the dual field theories~\cite{Calabrese:2004eu}.  Since the planar black brane metric~\eqref{blackBraneInf} is asymptotic AdS$_{d+1}$, the UV divergences take the universal form~\cite{Rangamani:2016dms,Hubeny:2012ry} 
\begin{align}
\label{AreaLogLaw}
    d> 2:\ \frac{\AreaLaw}{\epsilon^{d-2}} && {\rm or} && d=2:\ \log\frac{2}{\epsilon}
\end{align}
for some UV cutoff $\epsilon\ll 1$. When projecting the HRT surfaces to the boundary along constant infalling time to obtain the membranes, we subtract these UV divergences of the (H)RT surfaces. In the scaling limit, due to the $1/z^2$ factor in~\eqref{blackBraneInf}, the remaining parts of the (H)RT surfaces that compute finite entropy are inside or close to the horizon.\footnote{In the original membrane theory from holographic CFTs~\cite{Mezei:2018jco}, the prescription was to cut the HRT surfaces at the horizon and keep only the interior parts, as it is only these parts that grow linearly with time~\cite{Hartman:2013qma}. In a later work~\cite{Jiang:2024tdj}, it was discovered that for saturated RT surfaces, as well as certain half-space HRT surfaces with displacement in the $x$ direction, portions of the (H)RT surfaces that are outside but near the horizon can also contribute finite entropy in the scaling limit. As such, here we adopt the latter more general prescription to cut out the UV divergences of the (H)RT surfaces.}

\begin{figure}[htbp]
    \centering
    \includegraphics[width=.5\textwidth]{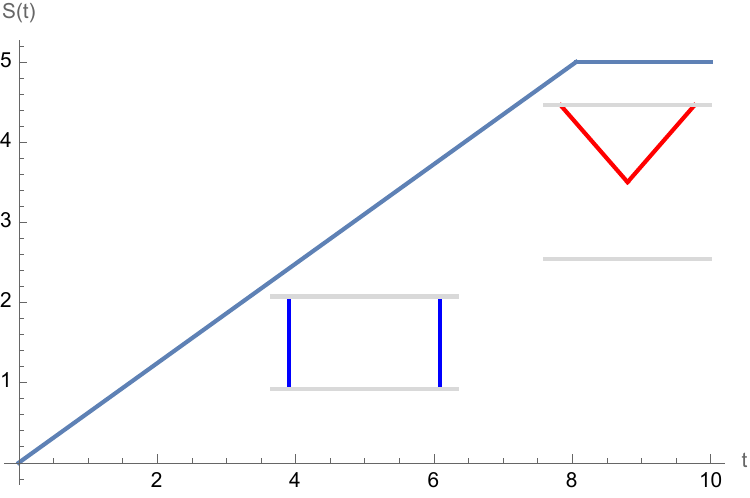}
    \qquad
    \caption{Time evolution of entanglement entropy of a single bounded interval in an unbounded system as described by the effective  membrane theory. The entanglement entropy first grows linearly and then saturates, and  is described in these two regimes by a vertical (blue) and cone (red) membranes, respectively. \label{fig:EEStrip}}
\end{figure}

For finite entangling regions, there is a candidate HRT surface which consists of two copies of the surface described above, one at each end of the subregion.  This leads to  the same linear growth of entanglement entropy. However, the entanglement entropy eventually saturates when the minimal area HRT surface instead becomes a connected surface outside the horizon. Projecting the latter to the boundary along constant infalling time gives a cone in the membrane description with the maximal slope $\pm v_B$~\cite{Jiang:2024tdj}.  The entanglement entropy from this surface is constant; the portion captured by the membrane picture is the extensive contribution, which is simply proportional to the size of the subregion, 
\begin{align}
    S=s_\text{th}  \ell \label{SatMembrane}
\end{align}
where $\ell$ is the size of the subregion in the boundary. The membrane picture is valid when $\ell$ is larger than all scales characterising the initial state. 
Thus for a single strip subregion, the entanglement entropy first grows linearly with slope determined by $v_E$ and then saturates to the thermal value for states dual to an AdS black brane (see Fig.~\ref{fig:EEStrip}). We have described how this behaviour arises from holography, but it is believed to apply more generally in any chaotic many-body system in the late time, large region regime where the effective membrane theory is valid (see e.g.~\cite{Nahum:2016muy,Jonay:2018yei,Zhou:2018myl,Zhou:2019pob,Rampp:2023vah,Vardhan:2024wxb}).

To simplify formulae, we rescale the boundary time  to absorb a factor of $v_E$, defining $t= v_E \, v$, so that the linear growth in Fig.~\ref{fig:EEStrip} is simply given by $S= 2s_\text{th} t$ (the factor of 2 appears because there are two vertical surfaces), and saturation occurs at $t = \frac{\ell}{2}$. 


\subsection{Membrane theory for reflected entropy}\label{SRreview}



To study the residual information and similar multi-party entanglement signals, we need an extension of the membrane theory describing the reflected entropy. This was obtained in \cite{Jiang:2024tdj}, where the authors studied dynamics of the reflected entropy in the holographic and effective membrane descriptions. In holography, the reflected entropy $S_{R}(A:B)$ is thought to be given by twice the area of the $AB$ entanglement wedge cross-section, as reviewed in section~\ref{sec:mphol}. This relation was established in the time-symmetric case \cite{Dutta:2019gen}, and is believed to be valid also in the time-dependent case. In \cite{Jiang:2024tdj}, a membrane theory for reflected entropy was proposed and checked with holography: the entanglement wedge cross-section for simple holographic systems was identified, and mapped to the boundary, showing agreement with the membrane effective description. As for the entanglement entropy, two qualitatively different kinds of membranes  appear: vertical membranes and diagonal membranes with slope $v_B$. 

\begin{figure}[htbp]
\centering
\includegraphics[width=.35\textwidth]{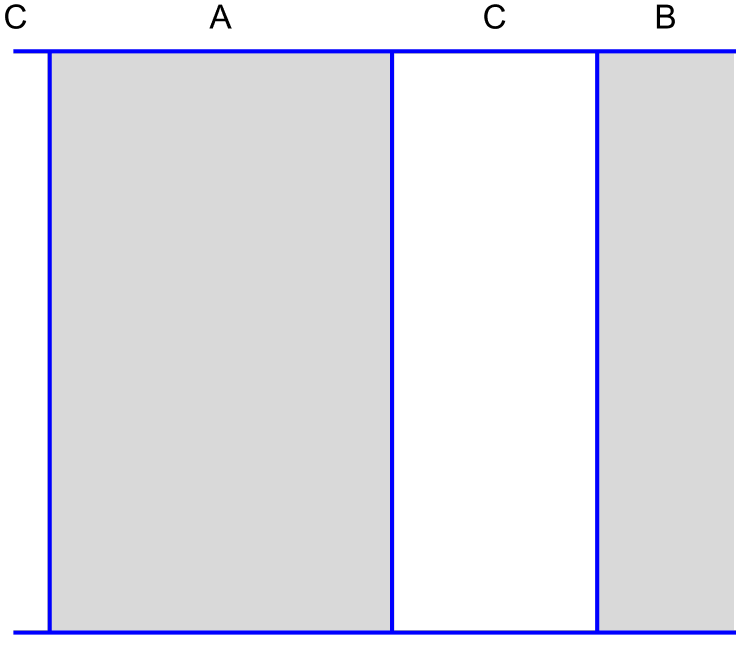}\hspace{5mm}
\includegraphics[width=.35\textwidth]{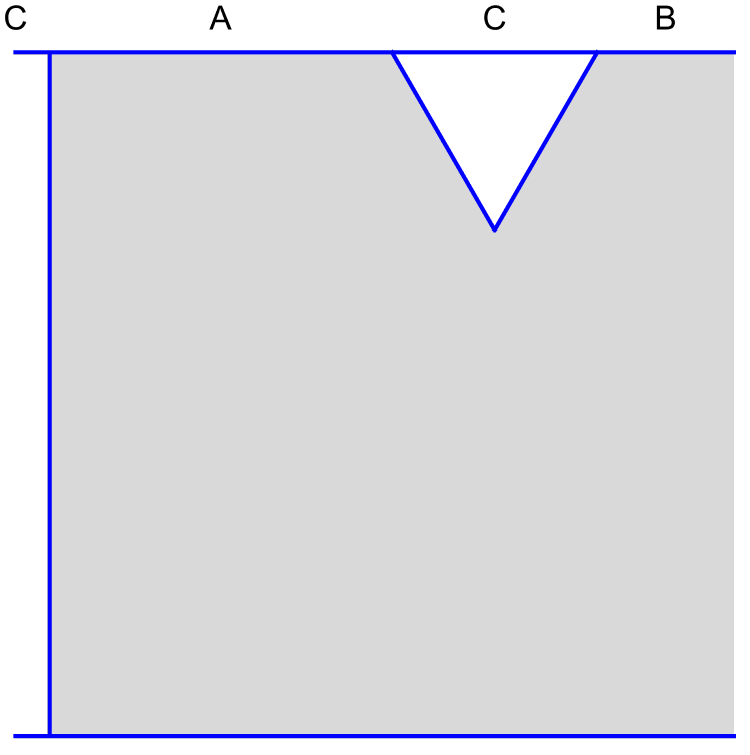}
\qquad
\caption{Membrane configurations computing $S(AB)$ when $B$ is semi-infinite, at $0< t<\frac{\ell_c}{2}$ (left) and $t>\frac{\ell_c}{2}$ (right). The $AB$ entanglement wedge is shaded. At $t=\frac{\ell_c}{2}$, the EW changes from disconnected to connected. \label{fig:SABSemiInf}}
\end{figure}

The authors of \cite{Jiang:2024tdj} work in a setting  with three regions, where $A$ is a finite interval,  $A = [0,\ell_A]$, $B$ is semi-infinite, $B = [\ell_A+\ell_C,\infty)$, and $C$ is the remainder of the system, which has two components, $C=(-\infty,0]\cup[\ell_A,\ell_A+\ell_C]$. For the $AB$ subregion, the minimal membranes computing the entanglement entropy are three vertical membranes dangling from the boundaries of $A$ and $B$ at early time, and a vertical membrane plus a cone with boundary slope $v_B$ spanning the finite part of $C$ at late times (see Fig.~\ref{fig:SABSemiInf}). In view of the entanglement wedge in the AdS bulk, in Fig.~\ref{fig:SABSemiInf} we refer to the region enclosed by the membranes, the $t=0$ slice, and the $t=T$ slice as the ``entanglement wedge'' in membrane theory. When the membrane theory describes a holographic CFT, this ``entanglement wedge'' can be obtained by projecting the bulk entanglement wedge to the boundary along constant infalling time. 

\begin{figure}[htbp]
\centering
\includegraphics[width=.35\textwidth]{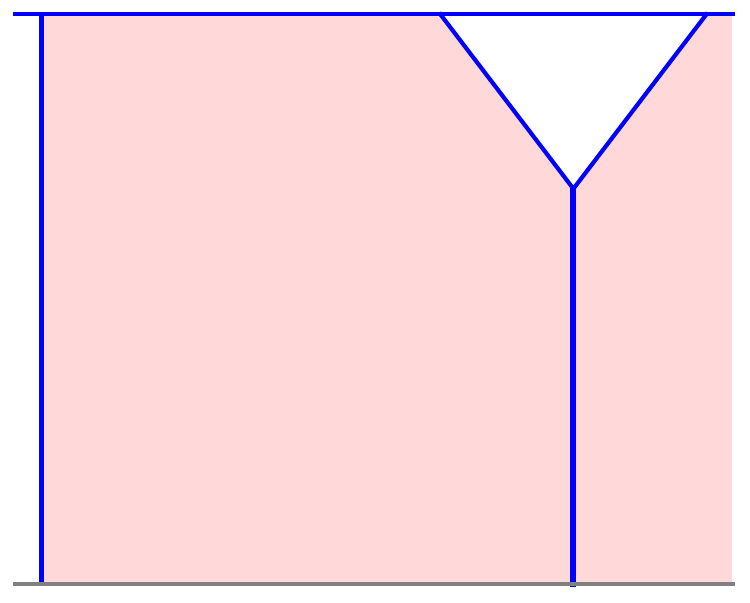}\hspace{5mm}
\includegraphics[width=.35\textwidth]{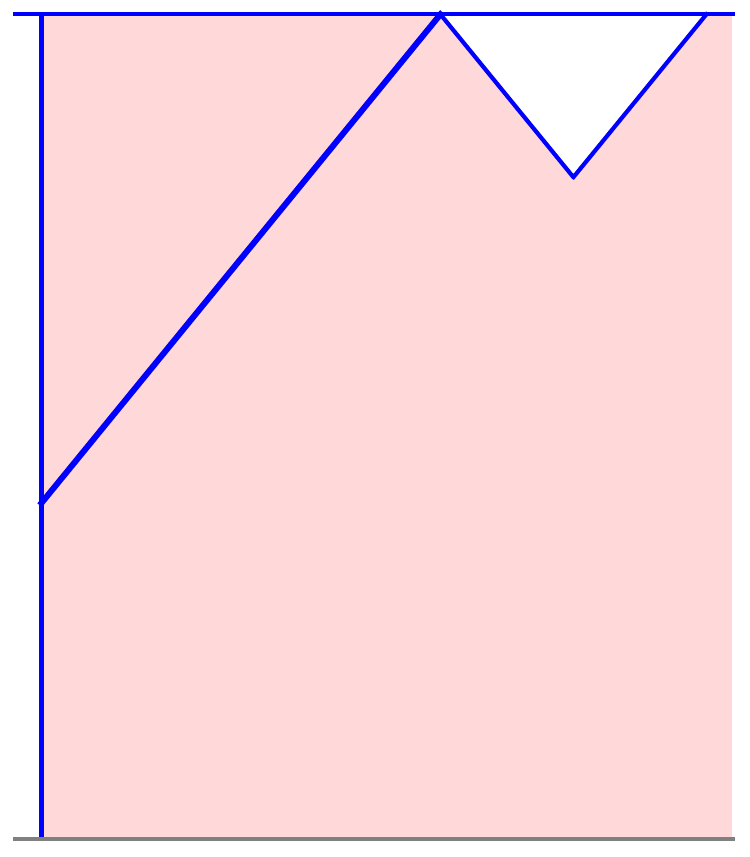}
\qquad
\caption{Minimal cross-section (thick) bisecting the $AB$ entanglement wedge (shaded region) at intermediate (left) and late (right) time. 
These entanglement wedge cross-sections compute the reflected entropy $S_R$. $t_0$ is the vertical distance between the top tip of the $v_B$ cone and the $t=T$ slice. \label{fig:MembraneEWCS}}
\end{figure}

The membrane effective theory for reflected entropy~\cite{Jiang:2024tdj} states that the reflected entropy in the scaling limit is computed by the minimal membrane bisecting the entanglement wedge. Again, inspired by entanglement wedge cross-section in holography, here we refer to this minimal membrane as the "entanglement wedge cross-section". When the $AB$ entanglement wedge is disconnected, as in the left panel of Fig.~\ref{fig:SABSemiInf}, the reflected entropy vanishes $S_{R}(A:B)=0$. 
When the $AB$ entanglement wedge becomes connected, as in the right panel in Fig.~\ref{fig:SABSemiInf}, there are two possibilities for the entanglement wedge cross-section, see Fig.~\ref{fig:MembraneEWCS}.  
At early time, the minimal cross-section is given by a vertical membrane hanging down from the tip of the $v_B$ cone, which gives linear growth for the reflected entropy, as in the earlier discussion of the entanglement entropy. However, the connected HRT surface for the finite part of $C$ hangs back to earlier times, so there is an offset $t_0$ relative to the formula for the entanglement entropy. 
This offset is determined by $v_B$, $2 v_B v_0 = \ell_C$, 
so $t_0 = \frac{v_E}{v_B} \frac{\ell_C}{2}$.  At later time, nevertheless, the minimal cross-section becomes the diagonal membrane that gives an extensive contribution to the reflected entropy proportional to the size of the boundary region, as in the $v_B$ cone for entanglement entropy. Thus 
\begin{equation}
\label{MembraneSR}
S_R(A:B) =2 s_\text{th}
\begin{cases}
0 \ &t < \frac{\ell_C}{2}\,, \\ 
t - \frac{v_E}{v_B} \frac{\ell_C}{2} \ \hspace{0.5cm} &\frac{\ell_C}{2} < t < \ell_A+\frac{v_E}{v_B} \frac{\ell_C}{2}\,, \\ 
\ell_A  \  &t > \ell_A+\frac{v_E}{v_B} \frac{\ell_C}{2}\,.
\end{cases}
\end{equation}
Notice there is a $\emph{discontinuous}$ jump in $S_R$ at the first transition where the entanglement wedge changes, although this becomes continuous if $v_E = v_B$.  When the membrane theory describes a
holographic CFT, it was checked in~\cite{Jiang:2024tdj} that the two types of minimal cross-section are indeed projections of the entanglement wedge cross-section in the dual gravity theory to the boundary along constant infalling time.

\subsection{Saturation of disjoint intervals}
\begin{figure}[htbp]
\centering
\includegraphics[width=.4\textwidth]{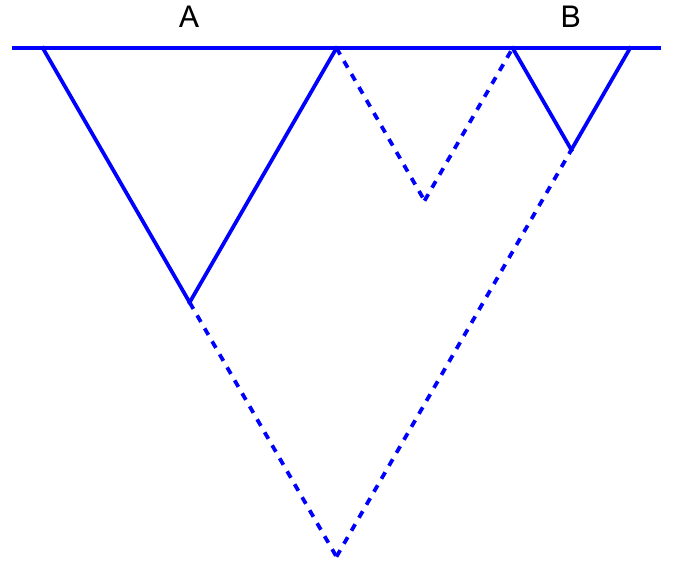}
\qquad
\caption{An example of different $v_B$ cone configurations for computing the saturation value of $S(AB)$. The solid lines correspond to the disconnected configurations, whereas the dashed lines represent the connected configurations. In the large subregion limit, the disconnected configurations always dominate. \label{fig:vBConesConnDis}}
\end{figure}

In what follows, we will be frequently dealing with entanglement dynamics of disjoint entangling subregions $A=\cup_i A_i$ ($i\geq 2$). For $n$ intervals, at later time there are multiple ways of connecting the boundaries of the intervals to form $v_B$ cones, leading to either $\emph{connected}$ or $\emph{disconnected}$ configurations, see Fig.~\ref{fig:vBConesConnDis}. For example, when $n=2$, from~\eqref{SatMembrane}, the connected and disconnected $v_B$ cone configurations compute EE
\begin{align}
S_{\text{conn.}}&=s_{\text{th}}\big((\ell_A+\ell_B+\ell_C)+\ell_C\big)\label{SConn}\\
S_{\text{dis.}}&=s_{\text{th}}(\ell_A+\ell_B)\label{SDis}
\end{align}
Hence, the disconnected $v_B$ cone configurations are $\emph{always}$ minimal. This is because the connected configurations overcount $\ell_C$ in between $A$ and $B$ twice. In fact, ~\eqref{SDis} is nothing more than another way of stating the $\emph{volume}$ law of EE for thermal states 
\begin{align}
    S=s_{\text{th}} \sum_i {\text{Vol}}(A_i)\label{volLaw}
\end{align}
which holds in the limit of large intervals. In our case with strip subregions, ${\text{Vol}}(A_i)=\ell_i$. The volume law indicates that the saturation pattern of forming disconnected $v_B$ cones remains true for subregions with arbitrary $n$ and generic sizes. 



For holographic CFTs in thermal states dual to asymptotically AdS black holes, if the entangling subregions are sufficiently large disjoint intervals, the above membrane theory discussion implies that the dual gravity theories always have $\emph{disconnected}$ entanglement wedges.   This is in contrast to the vacuum AdS cases, where the entanglement wedges can be either connected or disconnected depending on the separation of the two intervals~\cite{Headrick:2010zt}. The reason behind this difference is that entanglement in vacuum states obeys the $\emph{area}$ law~\cite{Bombelli:1986rw,Srednicki:1993im}. Holographically, the UV divergences of the RT surfaces~\cite{Rangamani:2016dms} always manifest this area law~\eqref{AreaLogLaw}. 


\section{Dynamics of $R_3$}
\label{sec:r3}

In this section we use the membrane effective description for the reflected entropy obtained in \cite{Jiang:2024tdj} to calculate the residual information in the simplest cases. We consider an infinite system with a division including one or two finite regions. We discuss here the ordinary membrane theory; the generalised membrane theory which describes the holographic calculations for $d=2$ has some quantitative differences, and is discussed in Section \ref{app:gen}.

\subsection{A single finite region}\label{SingleFiniteReg}

The simplest case to illustrate the dynamics of $R_3$ is where we divide the boundary into three regions along $x$: $A = (-\infty, 0)$,  $B = (0, \ell)$, and  $C= (\ell, \infty)$. 

\begin{figure}[htbp]
\centering
\includegraphics[width=.35\textwidth]{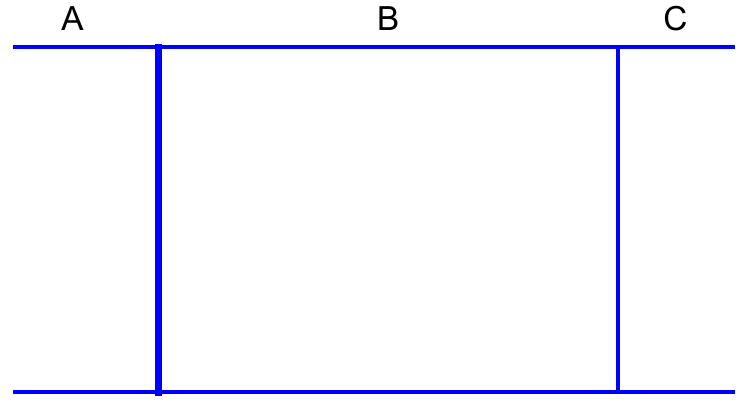}\hspace{5mm}
\includegraphics[width=.35\textwidth]{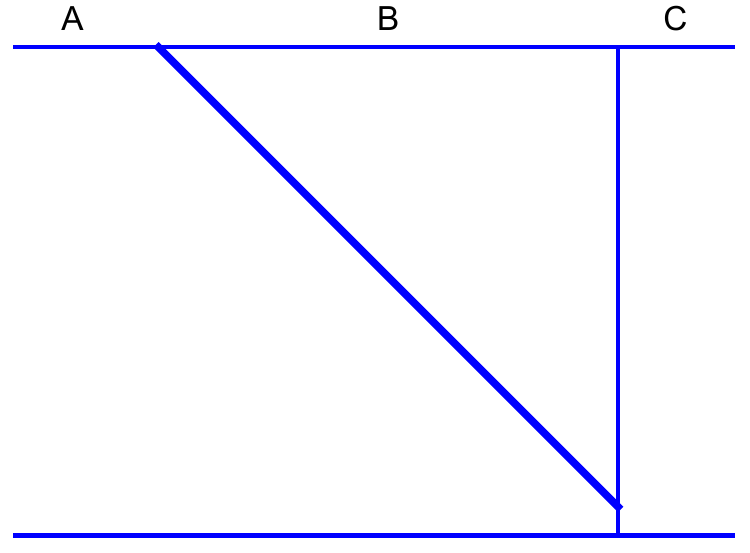}
\qquad
\caption{Membranes computing $S_R(A:B)$ (thick) in the setup of a single finite region $A = (-\infty, 0)$,  $B = (0, \ell)$, and  $C= (\ell, \infty)$.\label{fig:SRABC}}
\end{figure}

In this case, $S(A)$ and $S(C)$ grow linearly for all times, while $S(B)$ has linear growth followed by saturation. In this  case $S(A) = S(C)$ as they are both vertical surfaces. This gives

%
\begin{equation}
I(A:B) = S(A) + S(B) - S(C) = S(B) = s_{\text{th}}\label{IABSingleFiniteReg}
\begin{cases}
2  t &  t < \frac{\ell}{2} \,, \\
\ell &  t > \frac{\ell}{2} \,,  
\end{cases}
\end{equation}
where we have used the fact that for a pure state, $S(AB)=S(C)$. We also have
\begin{equation}
\label{IACSingleFiniteReg}
I(A:C) = S(A) + S(C) - S(B) = s_{\text{th}} 
\begin{cases}
0 &  t < \frac{\ell}{2} \,, \\
2 t - \ell &  t > \frac{\ell}{2} \,.
\end{cases}
\end{equation}
Since $A$, $B$ are adjacent regions, the $AB$ entanglement wedge cross-section is always non-empty. The cross-section is initially a vertical membrane hanging from the boundary, growing linearly in time. At $t=\ell$, this changes to a diagonal membrane extending across the $B$ region, giving a constant extensive answer for the entanglement wedge cross-section. See Figure~\ref{fig:SRABC}.\footnote{In the case of holographic CFTs, the saturated membrane on the right of Figure~\ref{fig:SRABC} is obtained from an Entanglement Wedge Cross-Section (EWCS) that skims through the planar black brane horizon and then ends on the HRT surface corresponding to the vertical membrane for S(C). This has been checked numerically in certain cases in~\cite{Jiang:2024tdj}. Notice that the diagonal membrane only captures the scaling behaviour of the EWCS and does not reflect its intersection with the HRT. } 
So we have\footnote{In the case of adjacent intervals, $S_R$ is not UV finite~\cite{Dutta:2019gen,Kudler-Flam:2020url}. In this case, we can  regularise the EWCS the same way as (H)RT, by subtracting UV divergences of the form~\eqref{AreaLogLaw}. These UV divergences cancel in $R_3$. \label{footnoteSR}}
\begin{equation}
\label{SRABSingReg}
S_R(A:B) = s_{\text{th}} 
\begin{cases}
 2 t &  t < \ell \,, \\
2\ell &  t > \ell \,.  
\end{cases}
\end{equation}
Hence 
\begin{equation}
\label{R3ABSingInt}
R_3(A:B) = S_R - I= s_{\text{th}} 
\begin{cases}
0 &  t < \frac{\ell}{2} \,, \\
2 t - \ell & \frac{\ell}{2} <  t < \ell \,, \\
\ell &  t > \ell \,.  
\end{cases}
\end{equation}
This exhibits the same qualitative dynamics as in the semi-infinite case considered in \cite{Jiang:2024tdj}; it is initially zero until the initial local entanglement has spread sufficiently to contribute to multiparty entanglement between the three regions, then it grows linearly for a period of time and then saturates. 

\begin{figure}[htbp]
\centering
\includegraphics[width=.35\textwidth]{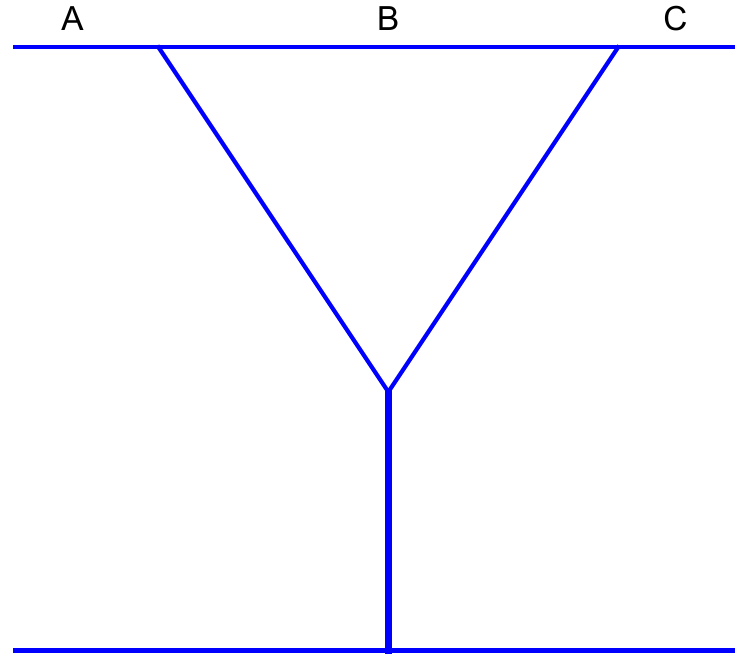}
\qquad
\caption{Membrane computing $S_R(A:C)$ (thick) in the setup of a single finite region.\label{fig:SRABCac}}
\end{figure}

By contrast, the $AC$ entanglement wedge is initially disconnected, only becoming connected when $S(B)$ saturates. Subsequently, the only possible cross-section is a linearly growing vertical surface, which hangs from the tip of the membrane describing $S(B)$, see Figure~\ref{fig:SRABCac}.

\begin{figure}[htbp]
\centering
\includegraphics[width=.6\textwidth]{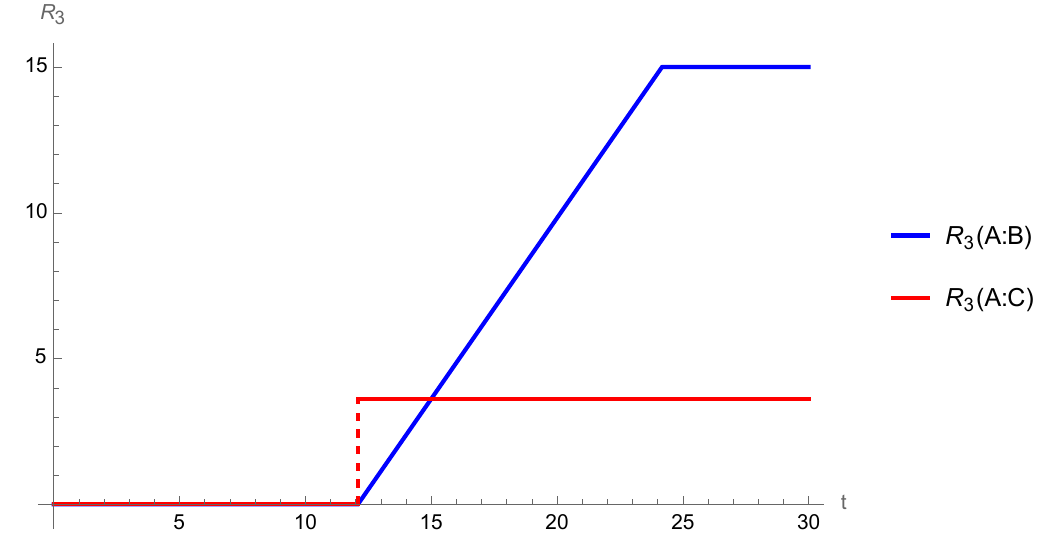}
\qquad
\caption{Time evolution of $R_3(A:B)$~\eqref{R3ABSingInt} (higher on the right, shown in blue) and $R_3(A:C)$~\eqref{R3ACSingInt} (lower on the right, shown in red) when $\ell=15$. \label{fig:R3SingInt}}
\end{figure}

This gives 
\begin{equation}
\label{SRACSingleFiniteReg}
S_R(A:C) = s_{\text{th}} 
\begin{cases}
0 &  t < \frac{\ell}{2} \,, \\
2 t-  \frac{v_E}{v_B} \ell &  t > \frac{\ell}{2} \, .
\end{cases}
\end{equation}
Note there is a discontinuity in $S_R$ at this transition, and we have an offset in time, as in the case discussed in \cite{Jiang:2024tdj} reviewed in~\eqref{MembraneSR}. As a result there is also a discontinuity in the residual information,
\begin{equation}
\label{R3ACSingInt}
R_3(A:C) = s_{\text{th}} 
\begin{cases}
0 &  t < \frac{\ell}{2} \,, \\
\ell \left( 1- \frac{v_E}{v_B} \right)  &  t > \frac{\ell}{2} \,.  
\end{cases}
\end{equation}
This illustrates the general point that different signals can have qualitatively different behaviour, see Figure~\ref{fig:R3SingInt}.

\subsection{Two finite regions}
\label{sec:r3twor}

Consider now a case where we divide the boundary into four pieces, that is with two finite intervals. To calculate $R_3$, we first need to combine two of the pieces to form a single region. There are two qualitatively different choices: to combine one of the semi-infinite pieces with a finite piece, or to combine the two semi-infinite pieces. The former is essentially what was considered in \cite{Jiang:2024tdj}, so we focus on the latter. That is, we consider dividing the boundary into three regions along $x$ which are now $A = (0,\ell_A)$, $B= (\ell_A, \ell_A + \ell_B)$, and $C = (-\infty,0) \cup (\ell_A + \ell_B,\infty)$. Let us suppose without loss of generality that $\ell_B < \ell_A$. We will consider both $R_3(A:B)$ and $R_3(A:C)$ ($R_3(B:C)$ is similar to $R_3(A:C)$). 

\begin{figure}[htbp]
\centering
\includegraphics[width=.35\textwidth]{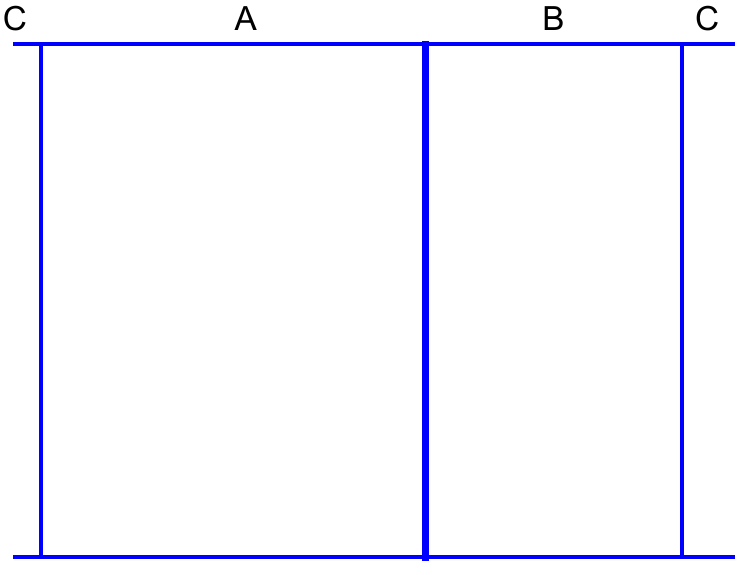}\hspace{5mm}
\includegraphics[width=.35\textwidth]{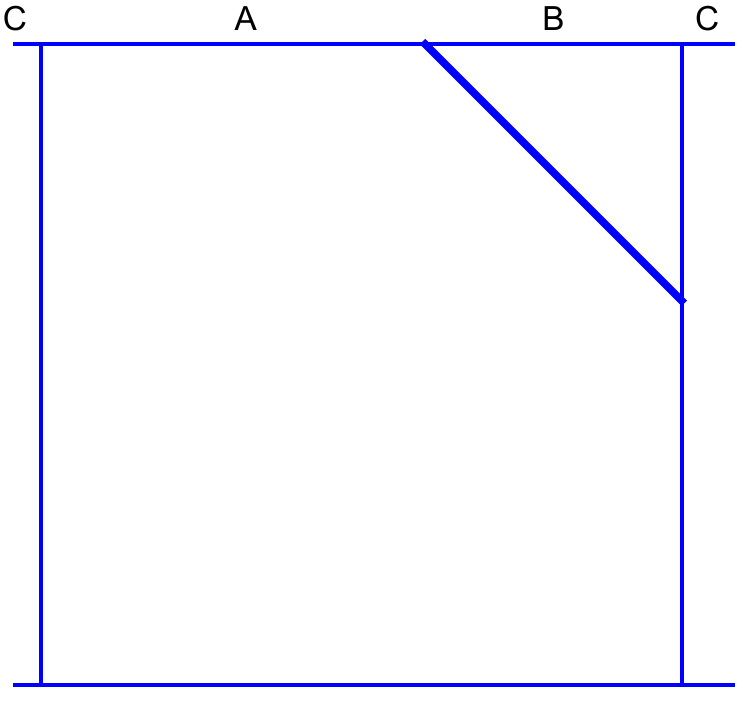}
\qquad
\caption{Membranes computing $S_R(A:B)$ (thick) in the setup of two finite regions $A = (0,\ell_A)$, $B= (\ell_A, \ell_A + \ell_B)$, and $C = (-\infty,0) \cup (\ell_A + \ell_B,\infty)$ ($\ell_A>\ell_B$). After the saturation of $S(AB)$ to a $v_B$ cone, we have $S_R(A:B)=0$, see Figure~\ref{fig:SatEWCSAdj}. \label{fig:SRCABC}}
\end{figure}

\begin{figure}[htbp]
\centering
\includegraphics[width=.5\textwidth]{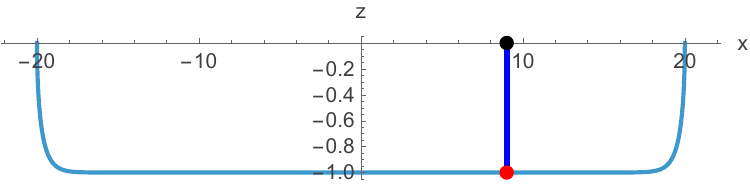}\vspace{0.5cm}
\includegraphics[width=.5\textwidth]{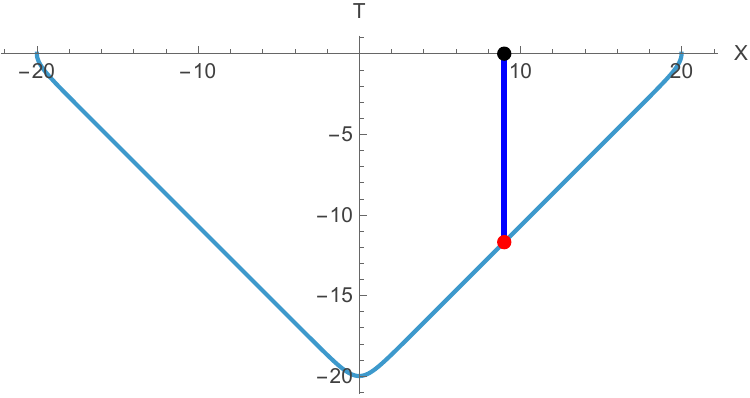}
\qquad
\caption{$\emph{Top}$: The minimal cross-section (blue) in finite adjacent intervals after saturation. It is almost radial as it needs to end perpendicularly  on the plateau of the RT surface perpendicularly, see Appendix~\ref{SatSRAdj}. $\emph{Bottom}:$ 
Projections of the minimal cross-section (blue) to the AdS boundary along constant infalling time. As discussed in appendix~\ref{SatSRAdj}, this vertical surface does not have a membrane description, and we have $S_R=0$ in this case contrary to what one might expect from the figure.  \label{fig:SatEWCSAdj}}
\end{figure}

The entropies $S(A)$, $S(B)$ and $S(AB) = S(C)$ exhibit the general linear growth followed by saturation behaviour of finite intervals. We thus have 
\begin{equation}
I(A:B) = s_{\text{th}} \label{IABTwoFiniteReg}
\begin{cases}
2  t &  t < \frac{\ell_B}{2} \,, \\
\ell_B & \frac{\ell_B}{2} <  t < \frac{\ell_A}{2} \,,   \\
\ell_A + \ell_B - 2  t & \frac{\ell_A}{2} <  t < \frac{\ell_A+\ell_B}{2} \,,  \\
0 &   t > \frac{\ell_A+ \ell_B}{2} \,.
\end{cases}
\end{equation}
The $AB$ entanglement wedge is always connected. The cross-section is initially a vertical surface hanging down from the boundary. It transitions to a surface extending across the $B$ region when $ t = \ell_B$, see Figure~\ref{fig:SRCABC}. When $S(AB)$ saturates at $ t = \ell_A + \ell_B$, the cross-section goes back to hanging down from the boundary but just to the $v_B$ cone membrane corresponding to $S(AB)$ (see Figure~\ref{fig:SatEWCSAdj}). However, this cross-section is no longer described by membrane theory. Intuitively, this is because it does not grow with $t$ or $x$ in the scaling limit. In Appendix~\ref{SatSRAdj}, we show using holography that for this cross-section $S_R=0$ in the scaling limit (after subtracting the UV divergences 
as discussed in footnote~\ref{footnoteSR}). So 

%
\begin{equation} \label{2srab}
S_R(A:B) = s_{\text{th}} 
\begin{cases}
2  t &  t < \ell_B \,, \\
2\ell_B & \ell_B <  t < \frac{\ell_A+\ell_B}{2} \,,  \\
0  &   t > \frac{\ell_A+ \ell_B}{2} \,.
\end{cases}
\end{equation}
The order of transitions in the residual information depends on whether $2 \ell_B$ is bigger or smaller than $\ell_A$. Suppose for definiteness $\ell_A < 2 \ell_B$,\footnote{The other case with $\ell_A>2\ell_B$ shows the same qualitative behaviour.} which includes the case where the two finite intervals are equal size. Then taking the difference between~\eqref{2srab} and~\eqref{IABTwoFiniteReg} we have 
\begin{equation}
\label{R3ABTwoReg1}
R_3(A:B) = s_{\text{th}} 
\begin{cases}
0 &  t < \frac{\ell_B}{2} \,, \\
2 t - \ell_B & \frac{\ell_B}{2} <  t < \frac{\ell_A}{2} \,,   \\
4 t - \ell_A - \ell_B & \frac{\ell_A}{2} <  t < \ell_B \,,   \\
2  t + \ell_B - \ell_A & \ell_B <  t < \frac{\ell_A+\ell_B}{2} \,,  \\
0  &   t > \frac{\ell_A+ \ell_B}{2} \,.
\end{cases}
\end{equation}
There is a single discontinuity in the residual information, at the final transition. At the transition point, the residual information reaches its maximum, $R_3(A:B) = 2\ell_B$, saturating a bound from \cite{Balasubramanian:2024ysu}, indicating that the extensive part of the entanglement of $B$ is entirely multipartite. 

\begin{figure}[htbp]
\centering
\includegraphics[width=.7\textwidth]{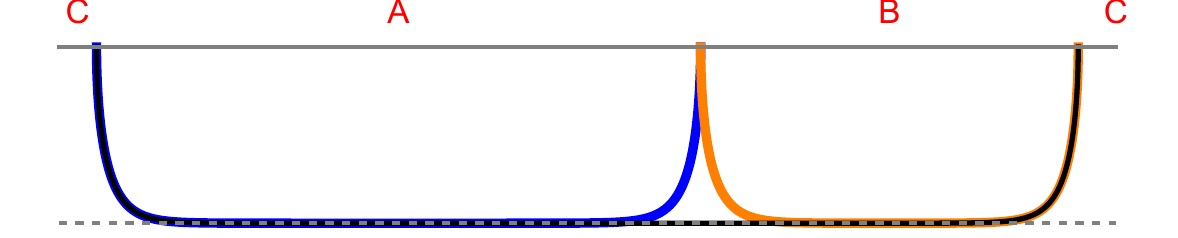}
\qquad
\caption{In the large subregion limit, RT surfaces have large plateau regions where they skim along the black hole horizon (dashed gray line), so the A RT surface (blue) and B RT surface (orange) lie close to the AB RT surface (black). \label{fig:RTClose}}
\end{figure}

Also, the jump is downward, so the residual information is not monotonically increasing. The residual information is only a signal of multiparty entanglement, so in general this does not mean that there is no multiparty entanglement in the final equilibrium state. But in holographic cases, in the final equilibrium here the $A$ and $B$ RT surfaces lie close to the $AB$ RT surface (see Figure~\ref{fig:RTClose}),  so it is reasonable to believe that the extensive entanglement here {\it is} purely bipartite, between $A$ and $C$ and $B$ and $C$.\footnote{The intuition here is that the entanglement can be modelled by a tensor network with locally entangled EPR pairs across the RT surface. This is a good model for the high-temperature thermofield double state, and was also shown to model hot multi-boundary wormholes~\cite{Marolf:2015vma}, which have a similar structure where the RT surfaces for different boundaries lie close to each other along most of their length. When the RT surfaces lie close to each other, the entangled EPR pairs across each surface simply join up to give local bipartite entanglement between the two sides.}

\begin{figure}[htbp]
\centering
\includegraphics[width=.3\textwidth]{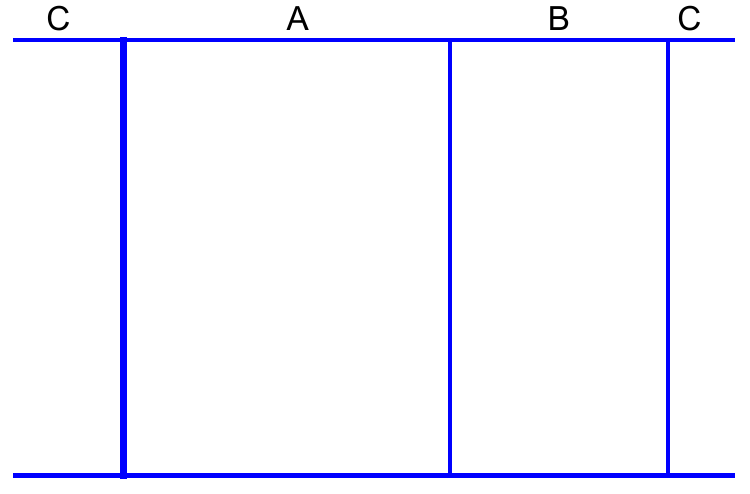}\hspace{3mm}
\includegraphics[width=.3\textwidth]{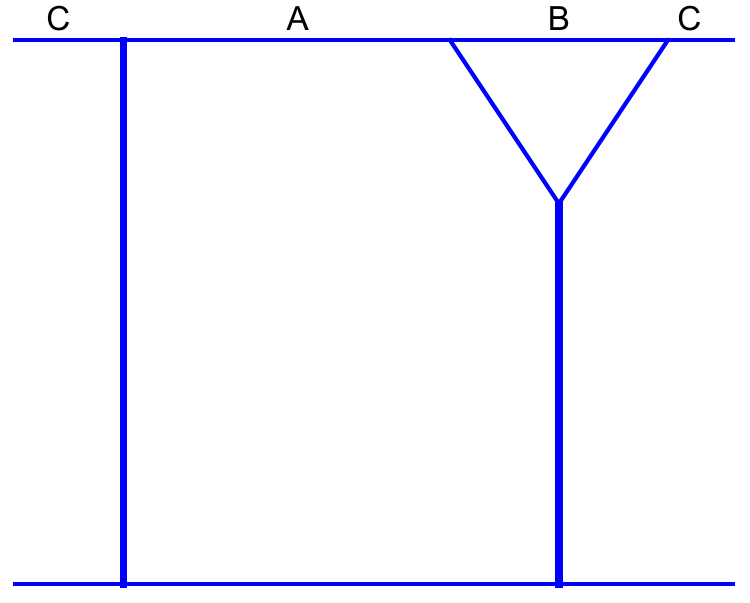}\hspace{3mm}
\includegraphics[width=.3\textwidth]{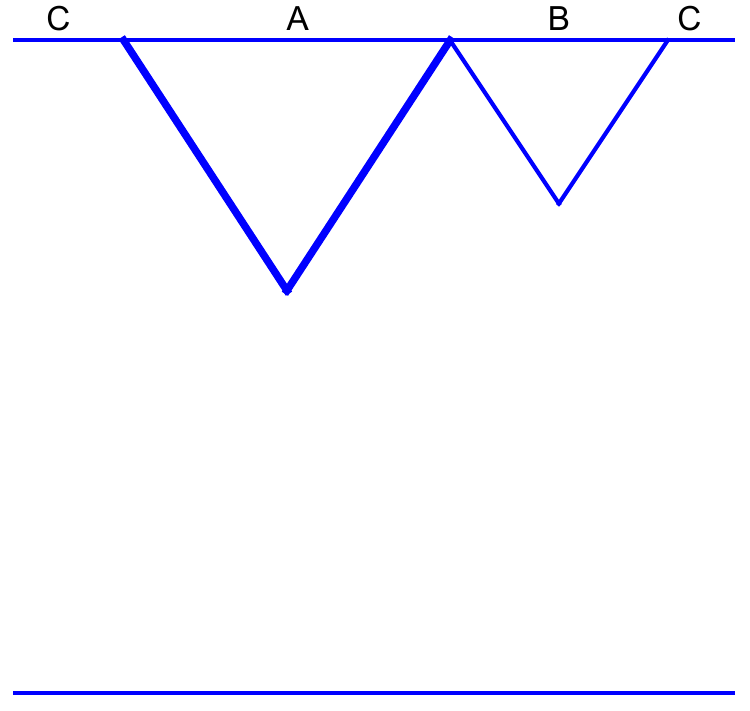}
\qquad
\caption{Membranes computing $S_R(A:C)$ (thick) in the setup of two finite regions. \label{fig:SRCABCac}}
\end{figure}

Let us also consider $R_3(A:C)$. We have
\begin{equation}
\label{IACTwoFiniteReg}
I(A:C) = s_{\text{th}} 
\begin{cases}
2 t &  t < \frac{\ell_B}{2} \,, \\
4 t - \ell_B & \frac{\ell_B}{2} <  t < \frac{\ell_A}{2} \,,   \\
2 t + \ell_A - \ell_B & \frac{\ell_A}{2} <  t < \frac{\ell_A+\ell_B}{2} \,,  \\
2 \ell_A &   t > \frac{\ell_A+ \ell_B}{2} \,.
\end{cases}
\end{equation}
The $AC$ entanglement wedge is always connected between $A$ and $C$. Initially the cross-section is just the vertical surface at $x=0$. When $B$ saturates, it acquires an additional component, a vertical surface hanging down from the $B$ membrane, then the two components join up. See Figure~\ref{fig:SRCABCac}.

Thus
\begin{equation} \label{2srac}
S_R(A:C) = s_{\text{th}} 
\begin{cases}
2 t &  t < \frac{\ell_B}{2} \,, \\
4 t - \frac{v_E}{v_B} \ell_B  & \frac{\ell_B}{2} < t < \frac{\ell_A}{2} + \frac{v_E}{v_B} \frac{\ell_B}{4} \,,   \\
2\ell_A &  t > \frac{\ell_A}{2} + \frac{v_E}{v_B} \frac{\ell_B}{4}  \,.
\end{cases}
\end{equation}
This gives
\begin{equation}
\label{R3ACTwoReg}
R_3(A:C) = s_{\text{th}} 
\begin{cases}
0 &  t < \frac{\ell_B}{2} \,, \\
\ell_B \left( 1 - \frac{v_E}{v_B} \right) & \frac{\ell_B}{2} <  t < \frac{\ell_A}{2} \,,   \\
2  t - \ell_A + \ell_B \left( 1 - \frac{v_E}{v_B} \right)  & \frac{\ell_A}{2} <  t < \frac{\ell_A}{2} + \frac{v_E}{v_B} \frac{\ell_B}{4} \,,  \\
\ell_A + \ell_B - 2 t & \frac{\ell_A}{2} + \frac{v_E}{v_B} \frac{\ell_B}{4}<  t < \frac{\ell_A+\ell_B}{2} \,,  \\
0 &   t > \frac{\ell_A+ \ell_B}{2} \,.
\end{cases}
\end{equation}
We see that the two residual information are non-zero in the same region, but their behaviour is otherwise quite different. $R_3(A:C)$ has an initial discontinuity, and then goes smoothly to zero at the end, so we see that $R_3$ can also decrease smoothly. It is also interesting to notice that $R_3(A:C)$ does not saturate the $2\ell_B$ bound. 


\begin{figure}[htbp]
\centering
\includegraphics[width=.6\textwidth]{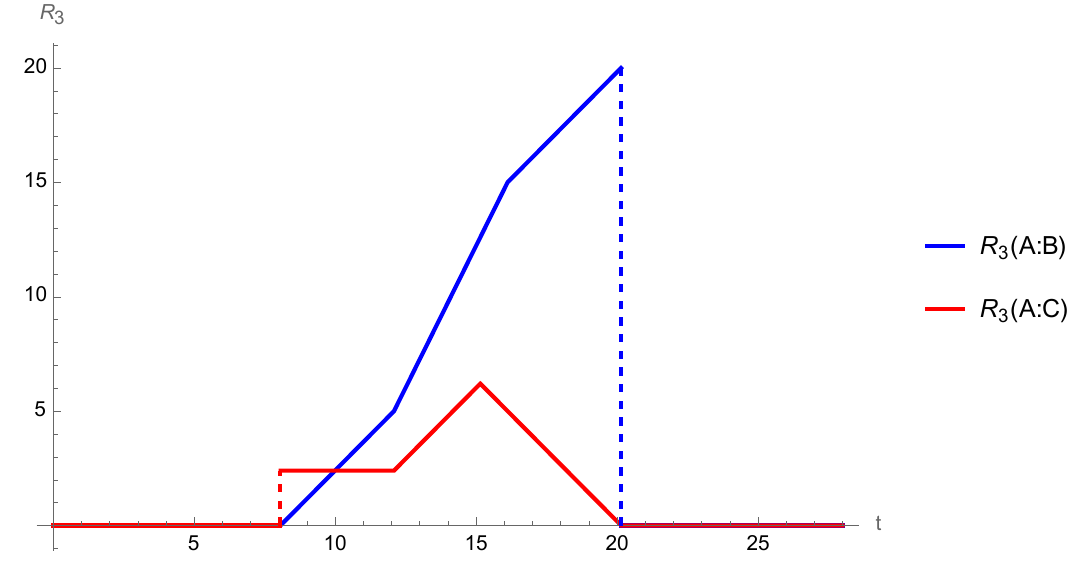}
\qquad
\caption{Time evolution of $R_3(A:B)$ (higher on the right, shown in blue) and $R_3(A:C)$ (lower on the right, shown in red) when $\ell_A=15$ and $\ell_B=10$. \label{fig:R3TwoReg}}
\end{figure}


\section{Dynamics of $I_3$ and $Q_4$}
\label{sec:I3}

We now turn to the dynamics of $I_3$ and $Q_4$. We will consider a single example, with regions $A = (0, \ell_A)$, $B = (\ell_A, \ell_A+\ell_B)$, $C= (\ell_A+\ell_B, \ell_A + \ell_B+\ell_C)$ and $D = (-\infty,0) \cup (\ell_A + \ell_B + \ell_C, \infty)$. For simplicity we restrict to $\ell_A = \ell_C$. The time-dependence of $I_3$ in this situation at finite time was previously considered in \cite{Balasubramanian:2011at}. We consider just the late-time, large region behaviour, in the context of the membrane theory. This captures the main qualitative features of the previous discussion. Our main interest here is to compare the behaviour of $I_3$ and the novel multiparty signal $Q_4$ introduced in \cite{Balasubramanian:2024ysu}. As we will see below, we find non-monotonicity\footnote{As already noted for $I_3$ in \cite{Balasubramanian:2011at}.} and saturation of bounds for both signals. For $I_3$ we cannot have discontinuities as it is just a combination of von Neumann entropies, but we do find discontinuities for $Q_4$.

If $\ell_A=\ell_C<\ell_B$, we find
\begin{equation}
\label{I3AdjacentNeqLenCase1}
I_3(A:B:C) = s_{\text{th}} 
\begin{cases}
0  &0<t < \frac{\ell_B}{2}\,, \\ 
\ell_B-2 t &\frac{\ell_B}{2}<t < \frac{\ell_A+\ell_B}{2}\,, \\ 
2 t-2\ell_A-\ell_B &\frac{\ell_A+\ell_B}{2} < t < \frac{2\ell_A+\ell_B}{2}\,, \\ 
0 &t>\frac{2\ell_A+\ell_B}{2}\,.
\end{cases}
\end{equation}
This gives a simple triangular profile for $-I_3$, initially increasing and then decreasing smoothly to zero.  


If  $\ell_B<\ell_A=\ell_C<2\ell_B$, we have a slightly more complex behaviour, 
\begin{equation}
\label{I3AdjacentNeqLenCase2}
I_3(A:B:C) = s_{\text{th}} 
\begin{cases}
0 \ &0<t < \frac{\ell_A}{2}\,, \\ 
2\ell_A-4 t\ &\frac{\ell_A}{2}<t < \frac{2\ell_A-\ell_B}{2}\,, \\ 
\ell_B-2 t\ &\frac{2\ell_A-\ell_B}{2}<t < \frac{\ell_A+\ell_B}{2}\,, \\ 
2 t-2\ell_A-\ell_B &\frac{\ell_A+\ell_B}{2} < t < \frac{2\ell_A+\ell_B}{2}\,, \\ 
0\ &t>\frac{2\ell_A+\ell_B}{2}\,.
\end{cases}
\end{equation}

\begin{figure}[htbp]
\centering
\includegraphics[width=.6\textwidth]{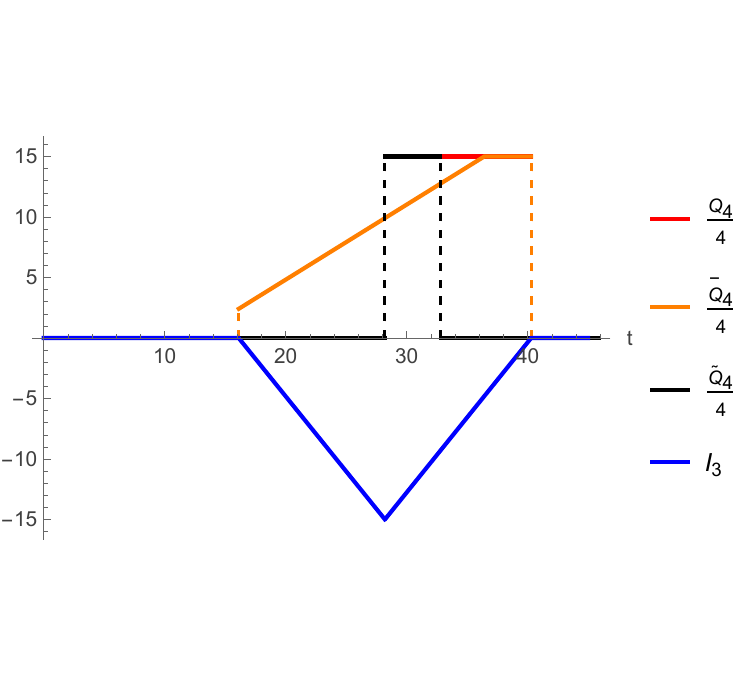}
\caption{Time evolution of $I_3(A:B:C)$~\eqref{I3AdjacentNeqLenCase1} (blue), $Q_4$~\eqref{Q4DABCD1} (red), $\Tilde{Q}_4$~\eqref{Q4T} (black) and $\bar{Q}_4$~\eqref{Q4B} (orange) for $\ell_A = \ell_C < \ell_B$. Here, we plot for $\ell_A=15$ and $\ell_B=20$. In this plot, the time at which $Q_4$ and $\Tilde{Q}_4$ start to be non-zero (the first black dashed vertical line) is the same; the time at which $Q_4$ and $\bar{Q}_4$ drop to 0 (orange dashed vertical line) is the same. \label{fig:I3Q41}}
\end{figure}
\begin{figure}[htbp]
\centering
\includegraphics[width=.6\textwidth]{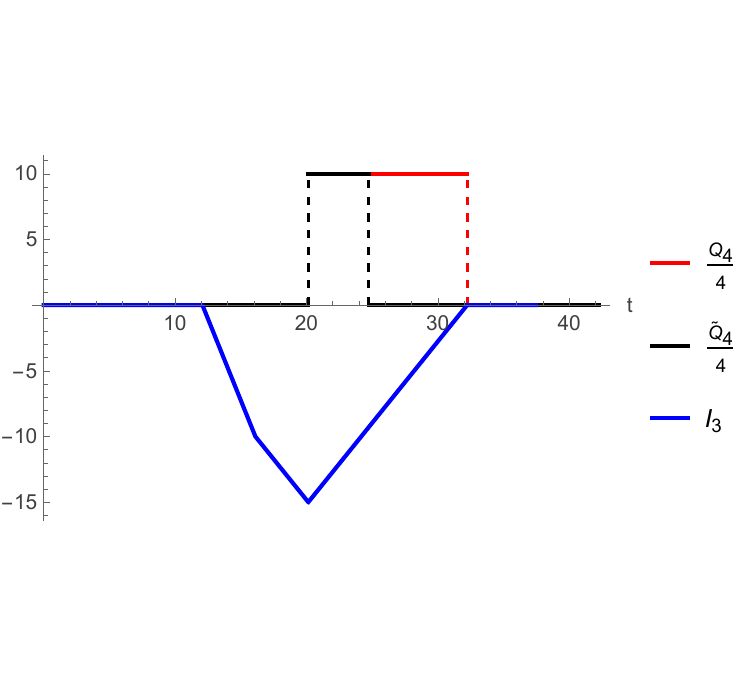}
\caption{Time evolution of $I_3(A:B:C)$~\eqref{I3AdjacentNeqLenCase1} (blue), $Q_4$~\eqref{Q4DABCD1} (red), and $\Tilde{Q}_4$~\eqref{Q4T} (black), for $\ell_B < \ell_A = \ell_C < 2 \ell_B$. Here, we plot for $\ell_A=15$ and $\ell_B=10$. In this plot, the time at which $Q_4$ and $\Tilde{Q}_4$ start to be non-zero (the first black dashed vertical line) is the same. \label{fig:I3Q42}}
\end{figure}
\begin{figure}[htbp]
\centering
\includegraphics[width=.6\textwidth]{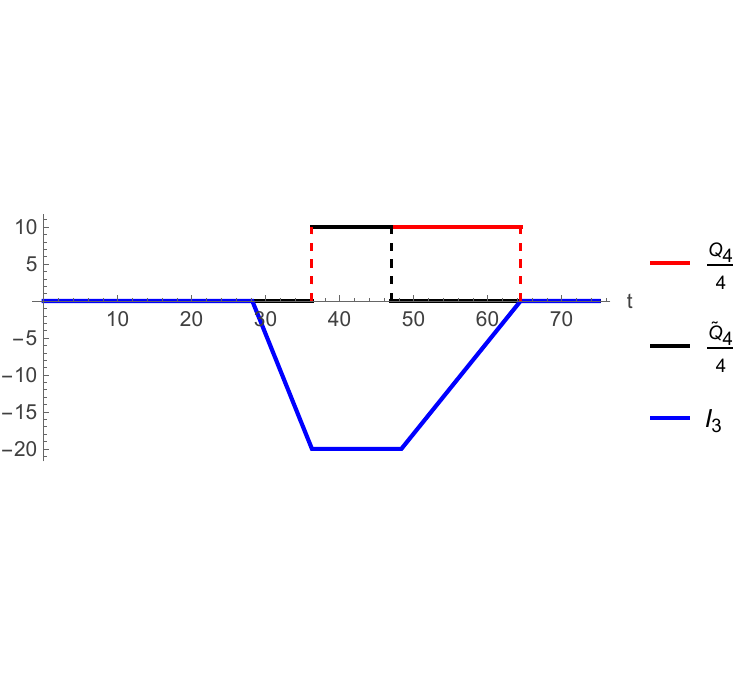}
\caption{Time evolution of $I_3(A:B:C)$~\eqref{I3AdjacentNeqLenCase1} (blue), $Q_4$~\eqref{Q4DABCD1} (red), and $\Tilde{Q}_4$~\eqref{Q4T} (black), for $2\ell_B < \ell_A = \ell_C$. Here, we plot for $\ell_A=35$ and $\ell_B=10$. \label{fig:I3Q43}}
\end{figure}

The most interesting case is when $\ell_A=\ell_C>2\ell_B$:
\begin{equation}
\label{I3AdjacentNeqLenCase3}
I_3(A:B:C) = s_{\text{th}} 
\begin{cases}
0 \ &0<t < \frac{\ell_A}{2}\,, \\ 
2\ell_A-4 t\ &\frac{\ell_A}{2}<t < \frac{\ell_A+\ell_B}{2}\,, \\ 
-2\ell_B\ &\frac{\ell_A+\ell_B}{2}<t < \frac{2\ell_A-\ell_B}{2}\,, \\ 
2 t-2\ell_A-\ell_B &\frac{2\ell_A-\ell_B}{2} < t < \frac{2\ell_A+\ell_B}{2}\,, \\ 
0\ &t>\frac{2\ell_A+\ell_B}{2}\,.
\end{cases}
\end{equation}
In this case there is a plateau region, where $I_3 = -2\ell_B$. This saturates a bound on $I_3$, indicating that the entanglement of $B$ is entirely four-partite. These results are plotted in in figures \ref{fig:I3Q41},
\ref{fig:I3Q42}, \ref{fig:I3Q43}.

\begin{figure}[htbp]
\centering
\includegraphics[width=.35\textwidth]{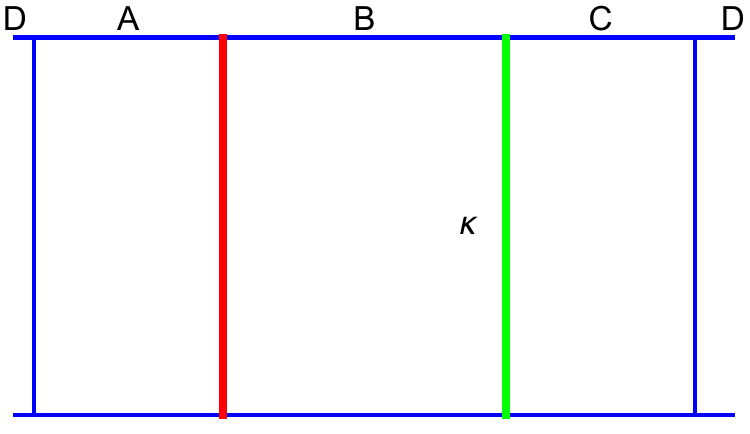}\hspace{3mm}
\includegraphics[width=.35\textwidth]{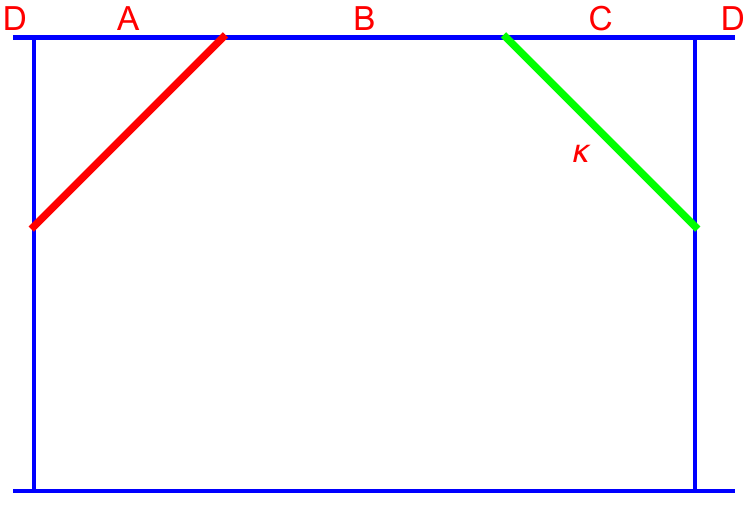}\vspace{6mm}
\includegraphics[width=.31\textwidth]{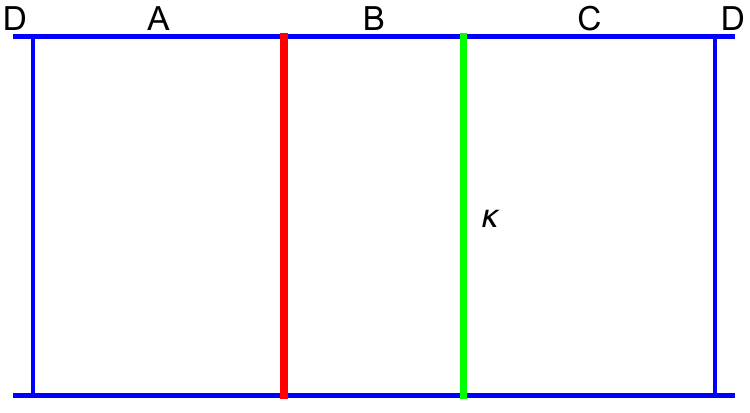}\hspace{2mm}
\includegraphics[width=.31\textwidth]{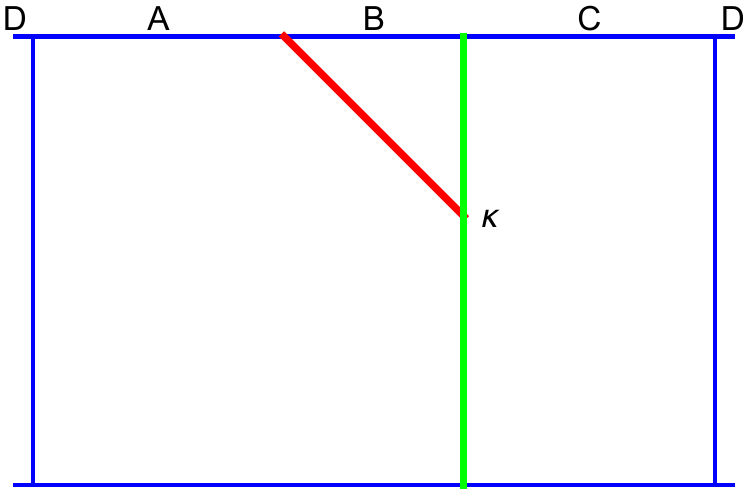}\hspace{2mm}
\includegraphics[width=.31\textwidth]{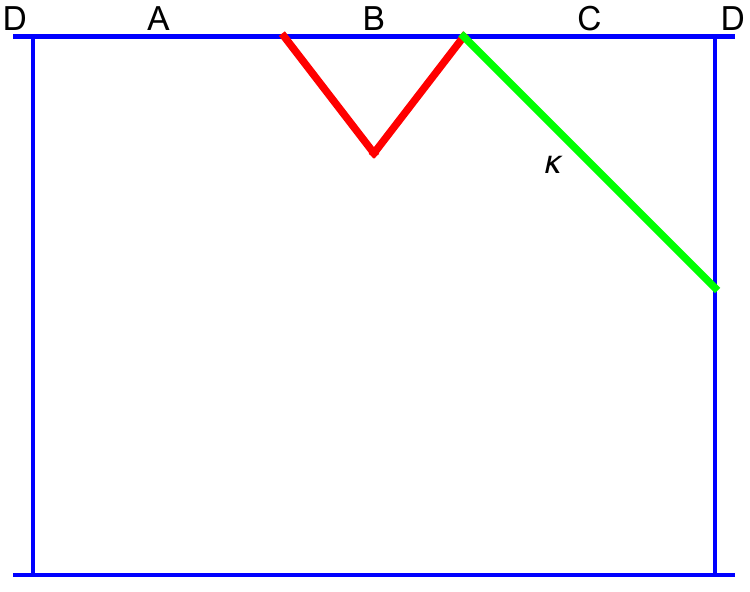}
\qquad
\caption{Minimal cross-sections computing $S_{AA^*A_*A^*_*}$ (red, thick) in the entanglement membrane description when $\ell_A<\ell_B$ (top row), and $\ell_A>\ell_B$ (bottom row). Notice that in the latter case, although the saturation of $\kappa$ (green, thick) changes the shape of the membrane computing $S_{AA^*A_*A^*_*}$, it does not change the value of $S_{AA^*A_*A^*_*}$ in the scaling limit. We only plot membranes before the saturation of the $ABC$ entanglement wedge, after which all the entanglement wedge cross-sections have vanishing area in the membrane description. \label{fig:kappa}}
\end{figure}

Next, we consider the calculation of $Q_4$ as in \cite{Balasubramanian:2024ysu}, tracing over $D$ first and then $CC^*$. 
The calculation of $S_R(A:B)$ in~\eqref{Q4} is similar to that in~\eqref{SRABSingReg}: we regard $CD$ as a single region, and the $AB$ entanglement wedge cross-section is then the same as those in Figure~\ref{fig:SRCABC}.\footnote{In Figure~\ref{fig:SRCABC} we considered only the case when $\ell_A>\ell_B$. Here we also allow $\ell_A<\ell_B$ and the $AB$ entanglement wedge cross-section saturates to the smaller value of $\ell_A$ and $\ell_B$.} To calculate $S_{AA^*A_*A^*_*}$, we consider the $AA^*BB^*$ entanglement wedge cross-section in the canonical purification of $\rho_{ABC}$. The first purification results in $EW(AA^*BB^*)$ bounded by a membrane $\kappa$ bisecting the EW of $AB$  (regarded as a single region) and $C$ (Figure~\ref{fig:kappa}). Notice that $\kappa$ itself possesses a non-trivial dynamic. After the second purification, $S_{AA^*A_*A^*_*}$ is computed by the minimal membrane bisecting $EW(AA^*BB^*)$ (Figure~\ref{fig:kappa}). 

By comparing the membranes computing $S_{AA^*A_*A^*_*}$ in Figure~\ref{fig:kappa} with those calculating $S_R(A:B)$ in Figure~\ref{fig:SRCABC}, we see that the EWCS for $S_{AA^*A_*A^*_*}$ and $S_R(A:B)$ at early times are the same.\footnote{In the case with $\ell_A > \ell_B$, the EWCS for $S_{AA^*A_*A^*_*}$ in the last panel in figure~\ref{fig:kappa} changes, but it has the same area.} Thus, the contributions cancel and $Q_4=0$. This ends when $S_{AB}$ saturates and we get figure~\ref{fig:SatEWCSAdj}. Then $S_R(A:B)=0$ in the membrane description, and $Q_4\neq 0$. This continues until $S_{ABC}$ similarly saturates, after which $S_{AA^*A_*A^*_*}$ also vanishes so we have $Q_4=0$ again. So $Q_4$ follows a unit-pulse-like time evolution profile

\begin{equation}
\label{Q4DABCD1}
Q_4 = s_{\text{th}} 
\begin{cases}
0 \ &0<t < \frac{\ell_A+\ell_B}{2}\,, \\ 
4 \text{min}\{\ell_A,\ell_B\} &\frac{\ell_A+\ell_B}{2}<t < \frac{\ell_A+\ell_B+\ell_C}{2}\,, \\ 
0\ &t>\frac{\ell_A+\ell_B+\ell_C}{2}\,.
\end{cases}
\end{equation}
where the minimisation in $Q_4$ between $\ell_A$ and $\ell_B$ captures how the saturated cross-section computing $S_{AA^*A_*A^*_*}$ exits the $AA^*BB^*$ entanglement wedge, see Figure~\ref{fig:kappa}. 

We plot the results when $\ell_A = \ell_C$ in Figures \ref{fig:I3Q41},
\ref{fig:I3Q42}, \ref{fig:I3Q43}. We see that the behaviour of $I_3$ and $Q_4$ are quite different; $I_3$ changes smoothly, while $Q_4$ is piecewise constant, and non-zero only in a subregion of the region where $I_3$ is varying. The discontinuities in $Q_4$ suggest phase transitions in the entanglement structure of the state, which are not evident from just considering $I_3$. In particular, $I_3$ goes smoothly to zero, but $Q_4$ shows that multiparty entanglement is {\it not} smoothly decreasing. It is still large at least up until $t = \frac{1}{2} (\ell_A + \ell_B + \ell_C)$, when it drops suddenly to zero, suggesting a sudden change of entanglement phase.

In Figure \ref{fig:I3Q43}, a bound from \cite{Balasubramanian:2024ysu} is saturated at the  plateau for $I_3$, indicating that the extensive entanglement of $B$ is purely four-party  at least in this period. In $Q_4$, the plateau also saturates a bound, which in this case indicates that the extensive entanglement of either $B$ (if $\ell_B < \ell_A$) or $A$ and $C$ (if $\ell_A = \ell_C < \ell_B$) is purely four-party  in that period. 

As in the calculation of $R_3$ in the previous section, the vanishing of $Q_4$ and $I_3$ at late times does not necessarily indicate the absence of four-party entanglement, but in the holographic cases, once all the finite regions have saturated the $ABC$ RT surface lies close to the individual $A$, $B$, $C$ RT surfaces, suggesting that the extensive part of the entanglement is entirely bipartite entanglement between each of the finite regions and $D$.

\begin{figure}[htbp]
\centering
\includegraphics[width=.4\textwidth]{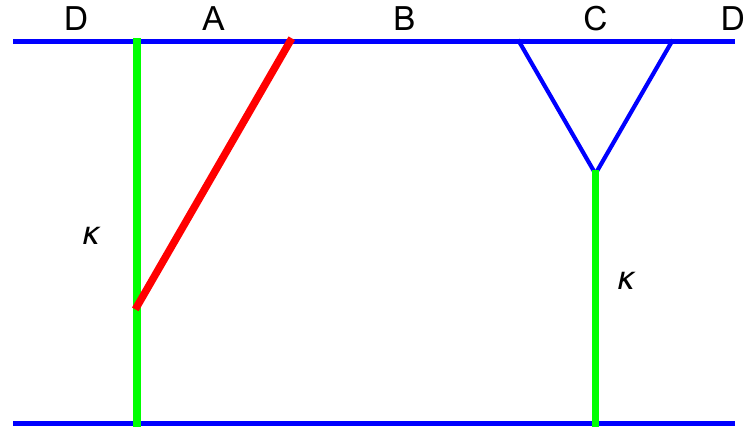}\hspace{4mm}
\includegraphics[width=.4\textwidth]{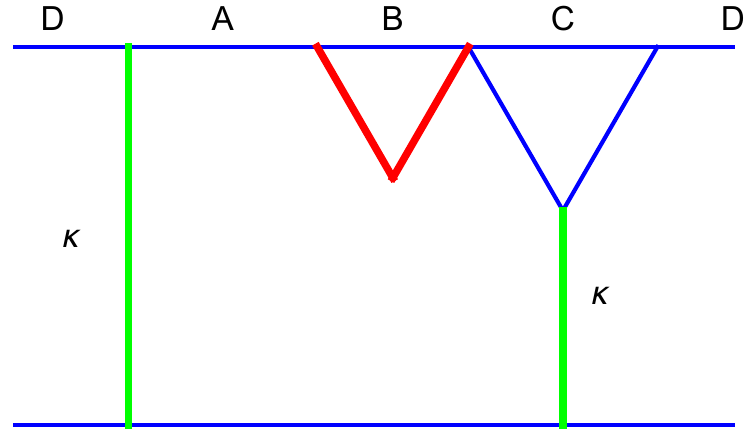}
\qquad
\caption{Examples of membranes computing $S_{AA^*A_*A^*_*}$ (red, thick) in $\tilde{Q}_4$. Notice that $\kappa$ (green, thick) behaves as $S_R(A:C)$ by regarding $AB$ as a single interval; $S_{AA^*A_*A^*_*}=0$ after $\kappa$ saturates. \label{fig:Q4T}}
\end{figure}

In obtaining~\eqref{Q4DABCD1}, we have traced out $D$ first and then $CC^*$. We can also consider tracing out $C$ first and the $DD^*$. This gives an independent signal, which we will denote as $\Tilde{Q}_4$. When tracing out $C$ in the first purification, the entanglement wedge of $AB$ (regarded as a single interval) and $D$ is bipartitioned by minimal membranes $\kappa$ that behave as $S_R(A:C)$ in Figure~\ref{fig:SRCABCac}. $S_{AA^*A_*A^*_*}$ in the second purification is then computed by the minimal cross-section that bisects the $AA^*BB^*$ entanglement wedge bounded by $\kappa$, see Figure~\ref{fig:Q4T}. As such, when fixing $\ell_A=\ell_C$, we have 
\begin{equation}
\label{Q4T}
\Tilde{Q}_4 = s_{\text{th}} 
\begin{cases}
0 \ &0<t < \frac{\ell_A+\ell_B}{2}\,, \\ 
4 \text{min}\{\ell_A,\ell_B\} &\frac{\ell_A+\ell_B}{2}<t < \frac{\ell_A+\ell_B}{2}+\frac{v_E}{4v_B}\ell_A\,, \\ 
0\ &t>\frac{\ell_A+\ell_B}{2}+\frac{v_E}{4v_B}\ell_A\,.
\end{cases}
\end{equation}
This is also plotted in figures~\ref{fig:I3Q41},
\ref{fig:I3Q42}, \ref{fig:I3Q43}. Note that compared to $Q_4$, the duration for $\Tilde{Q}_4$ to be non-zero is shorter. This illustrates the advantage of having multiple signals, as $Q_4$ and $I_3$ capture the existence of 4-party entanglement that $\Tilde{Q}_4$ fails to detect after its disappearance. 

\begin{figure}[htbp]
\centering
\includegraphics[width=.45\textwidth]{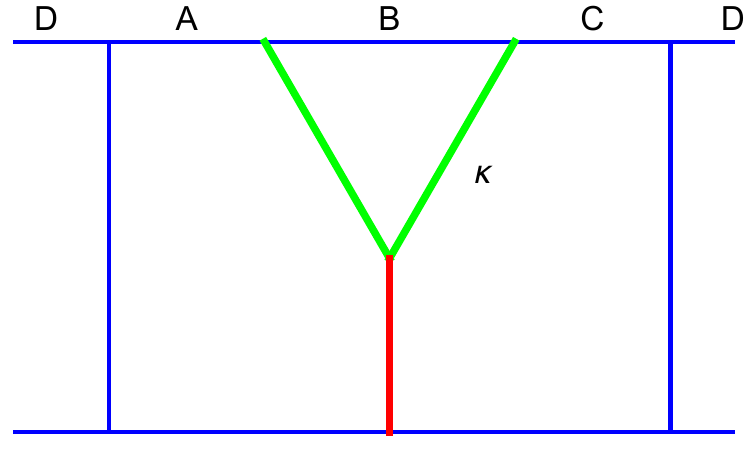}
\qquad
\caption{An example of the membrane computing $S_R(A:C)$ or $S_{AA^*A_*A^*_*}$ in $\bar{Q}_4$, depicted in red. In the former case, $\kappa$ along with the two outermost vertical membranes together are the membranes for $S(AC)$. \label{fig:Q4B}}
\end{figure}

One can also consider tracing over different regions in the definition of $Q_4$. For example, we can trace out $D$ and $B$ to obtain another version of $Q_4$, which we denote as $\bar{Q}_4$. When $\ell_A>\ell_B$, the membranes for $S_R(A:C)$ consist of a vertical membrane dangling from the tip of the $v_B$ cone anchored to $B$ at early time as shown in Figure~\ref{fig:Q4B}, followed by a diagonal membrane of slope $|v_B|$ exiting the $AC$ entangling wedge through either of its boundaries. When $\ell_A<\ell_B$, we simply have $S_R(A:C)=0$ at all time, as the $AC$ entanglement wedge is always disconnected. 

To calculate $S_{AA^*A_*A^*_*}$, let us first consider the minimal cross-section $\kappa$ that bisects the entanglement wedge of $AC$ (regarded as one disjoint subregion) and $B$ in the first purification. At early time, $\kappa$ consists of two vertical membranes. These will later be replaced by  a single triangle anchored to $B$ (see Figure~\ref{fig:Q4B}) or two diagonal membranes with slope $\pm v_B$ exiting the $ABC$ entanglement wedge, depending on whether $\ell_B<2\ell_A$ or $\ell_B>2\ell_A$. In the latter case, the $AA^*CC^*$ entanglement wedge is always disconnected, and $S_{AA^*A_*A^*_*}=0$ at all time. As for the former, $S_{AA^*A_*A^*_*}$ is computed by the minimal cross-section bi-partitioning the $AA^*CC^*$ entanglement wedge bounded by $\kappa$, which is a vertical membrane dangling from
the tip of the $v_B$ cone at early time (as  shown in Figure~\ref{fig:Q4B}) and a diagonal membrane of slope $|v_B|$ exiting the $AA^*CC^*$ EW at late time. Notice that these are the $\emph{same}$ membranes as those computing $S_R(A:C)$. Hence, $\bar{Q}_4\neq 0$ only when $\ell_A<\ell_B<2\ell_A$: for this choice of $\ell_A$ and $\ell_B$, $S_R(A:C)=0$ at all times, but $S_{AA^*A_*A^*_*}$ undergoes a non-trivial dynamic. When $\ell_B>\ell_A$, both $S_R(A:C)$ and $S_{AA^*A_*A^*_*}$ are $0$ all the time; when $\ell_A>\ell_B$, both $S_R(A:C)$ and $S_{AA^*A_*A^*_*}$ have non-trivial dynamics, but they are computed by the same membranes at any time, so their contributions cancel. As such, we have that for $\ell_A<\ell_B<2\ell_A$,
\begin{equation} \label{Q4B}
\bar{Q}_4 = s_{\text{th}} 
\begin{cases}
0 &  t < \frac{\ell_B}{2} \,, \\
4 \big(t - \frac{v_E}{2 v_B} \ell_B\big)  &  \frac{\ell_B}{2} < t < \ell_A + \frac{v_E}{2v_B} \ell_B \,, \\
4 \ell_A & \ell_A + \frac{v_E}{2v_B} \ell_B < t < \ell_A + \frac{\ell_B}{2}  \,,   \\
0 &  t > \ell_A + \frac{\ell_B}{2} \,.
\end{cases}
\end{equation}
This is plotted in Figure~\ref{fig:I3Q41}. In the cases in Figures~\ref{fig:I3Q42} and ~\ref{fig:I3Q43}, $\bar Q_4$ is always zero.

Unlike $Q_4$ or $\Tilde{Q}_4$, $\bar{Q}_4$ does not display the dynamics of a unit pulse. Moreover, $\bar{Q}_4$ is non-zero in the whole interval where $I_3$ is nonvanishing. We also note (see Figure~\ref{fig:I3Q41}) that the times at which $Q_4$, $\Tilde{Q}_4$ and $\bar{Q}_4$ saturate the $4\ell_A$ bound are different. We believe that the time at which the extensive part of the entanglement of $A$ is entirely multipartite is when $Q_4$ stays at its bound $Q_4= 4\ell_A$, i.e. when $\frac{\ell_A+\ell_B}{2}<t < \frac{\ell_A+\ell_B+\ell_C}{2}$. During this time $\Tilde{Q}_4$ and $\bar{Q}_4$ could be 0 or non-maximal, which means that they fail to reflect the fact that the entanglement of $A$ is entirely multipartite. The above different behaviours among $Q_4$, $\Tilde{Q}_4$, and $\bar{Q}_4$ again exemplify the advantage of having multiple signals, as each of them only captures certain aspects of 4-partite entanglement.


There are a number of other versions of $Q_4$: these can be either positive or negative between $\pm 4{\rm min}\{\ell_A,\ell_B\}$, and might saturate these bounds or not, depending on the subregion sizes. 
%



\section{Higher partite entanglement signals}
\label{sec:higher}

We can extend the above calculations to higher numbers of parties. The extension is conceptually straightforward, and does not seem to lead to qualitatively new features, but the calculations become more complicated due to the increasing number of regions. 

Consider for example the five-party entanglement signal $R_5$. Suppose we examine a system with subregions $A=(0,\ell)$, $B=(\ell,2\ell)$, $C=(2\ell,3\ell)$, $D=(3\ell,4\ell)$, $E=(-\infty,0)\cup (4\ell,+\infty)$, and trace over $E$. We have
\begin{equation}
R_5(A:B:C:D) = \frac{1}{2} I_4(AA^*:BB^*:CC^*:DD^*) - I_4(A:B:C:D),
\end{equation}
where
\begin{equation}
\label{R51}
    \begin{split}
   \frac{1}{2} I_4(AA^*:BB^*:CC^*:DD^*) =& I_3(AA^*:BB^*:CC^*) \\ =& S(AA^*) + S(BB^*)+ S(CC^*)+S(DD^*)\\ & - (S(AA^*BB^*) + S(AA^*CC^*) + S(BB^*CC^*)),     
    \end{split}
\end{equation}
and 
\begin{equation}
    \begin{split}
  I_4(A:B:C:D) =& S(A) + S(B)+ S(C)+ S(D) \\ &- (S(AB)+S(AC)+S(AD) + S(BC) + S(BD)+ S(CD) ) \\ &+ S(ABC) + S(ABD) + S(ACD) + S(BCD) - S(ABCD).          
    \end{split}
\end{equation}
%

\begin{figure}[htbp]
\centering
\includegraphics[width=.35\textwidth]{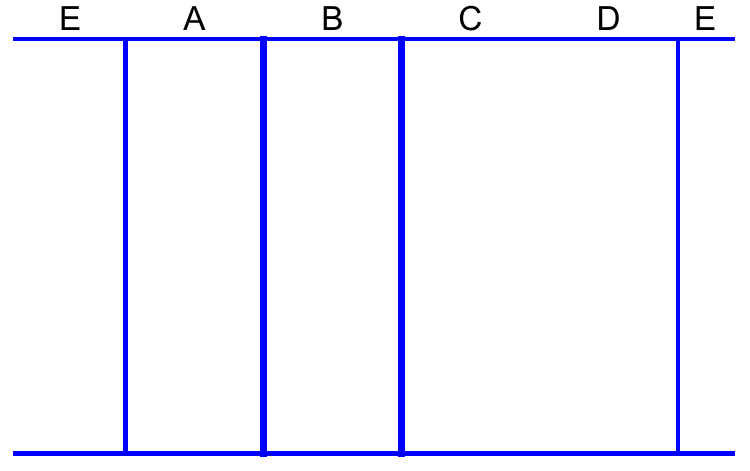}\hspace{3mm}
\includegraphics[width=.35\textwidth]{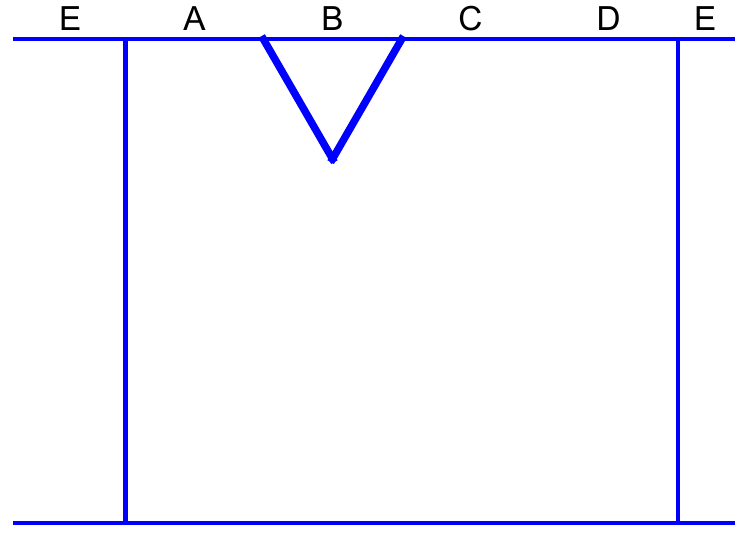}\vspace{5mm}
\includegraphics[width=.31\textwidth]{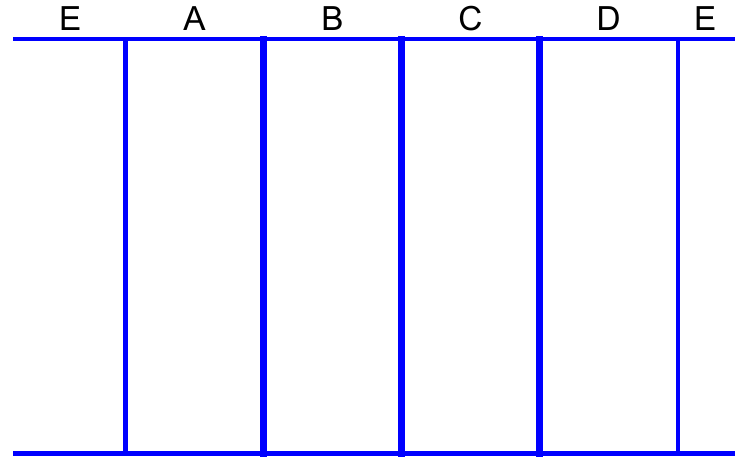}\hspace{3mm}
\includegraphics[width=.31\textwidth]{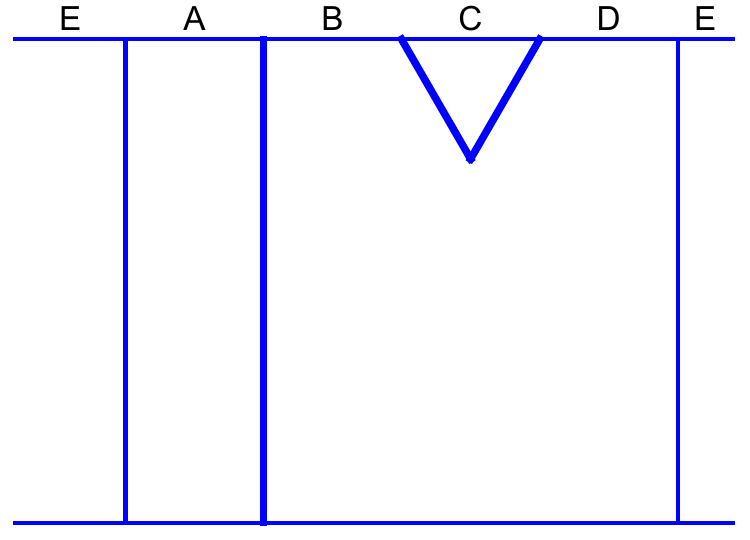}\hspace{3mm}
\includegraphics[width=.31\textwidth]{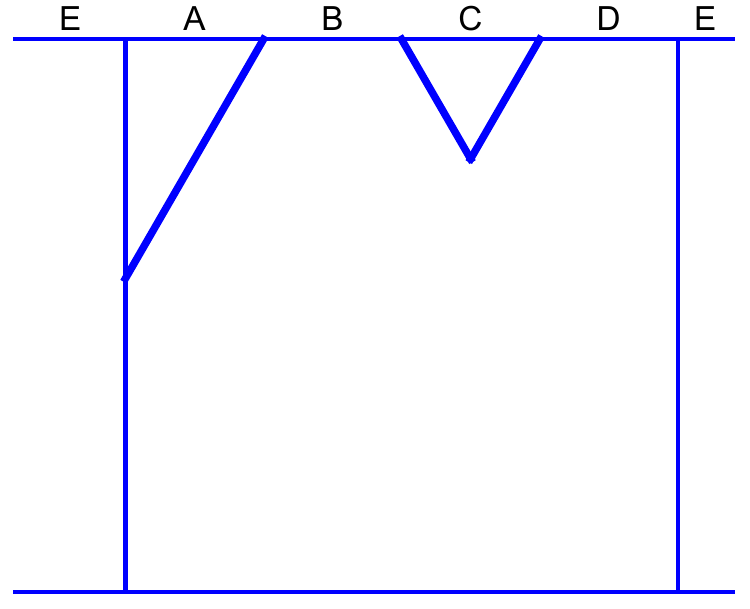}
\qquad
\caption{Membranes computing $S(BB^*)$ (top row) and $S(AA^*CC^*)$ (bottom row) in $R_5(A:B:C:D)$. We only plotted membranes before the saturation of the $ABCD$ EW, after which all the entanglement wedge cross-sections have vanishing area in the membrane description. \label{fig:R5}}
\end{figure}

We have $S(A) = S(B) = S(C) = S(D) = S_1$, $S(AB) = S(BC) = S(CD) = S_2$, $S(AC) = S(AD) = S(BD) = 2 S_1$, $S(ABC) = S(BCD) = S_3$, $S(ABD) = S(ACD) = S_2+S_1$, and $S(ABCD) = S_4$, where we use $S_n$ to denote the entanglement entropy of an interval of length $n \ell$. This gives
\begin{equation}
  I_4(A:B:C:D) = 2 S_3 - S_2 - S_4 = s_\mathrm{th}  \begin{cases} 0 & t < \ell , \\ 
  2t - 2\ell & \ell < t < \frac{3\ell}{2}, \\ 
  4 \ell-2t & \frac{3\ell}{2} < 2 \ell, \\
  0 & t > 2\ell. \end{cases}    
\end{equation}
The $ABCD$ entanglement wedge is always connected. Before the transition at $t = 2\ell$, it is bounded by two vertical surfaces (Figure~\ref{fig:R5}). The purification-based quantities $S(XX^*)$ in~\eqref{R51} are computed by the minimal cross-sections that bisect the entanglement wedge of $X$ and $ABCD/X$ (see Figure~\ref{fig:R5} for examples). As such, we have $S(AA^*) = S(DD^*) = S_2$, $S(BB^*) = S(CC^*) = 2 S_1$, $S(AA^*BB^*) = S_4$, $S(AA^*CC^*) = S_2 + 2 S_1$, and $S(BB^*CC^*) = 2S_2$. After the transition, the entanglement wedge only extends down to the horizon, and all the canonical purification quantities are given by vertical surfaces which extend down to the horizon (Figure~\ref{fig:SatEWCSAdj}), so $I_4(AA^*:BB^*:CC^*:DD^*) = 0$ for $t > 2\ell$. Thus
\begin{equation}
\frac{1}{2}   I_4(AA^*:BB^*:CC^*:DD^*) = s_\mathrm{th}  \begin{cases} 0 & t < \frac{\ell}{2} , \\ 
  2\ell - 4t & \frac{\ell}{2} < t < \ell, \\ 
  -2t & \ell < 2 \ell, \\ 
  0 & t > 2\ell. \end{cases}    
\end{equation}

This finally gives
\begin{equation}
\label{R5EABCDE}
R_5(A:B:C:D) = s_{\text{th}} 
\begin{cases}
0  &0<t < \frac{\ell}{2}\,, \\ 
2\ell-4 t &\frac{\ell}{2}<t < \frac{3\ell}{2}\,, \\ 
-4\ell &\frac{3\ell}{2}<t <2\ell\,, \\ 
0&t>2\ell\,.
\end{cases}
\end{equation}

\begin{figure}[htbp]
\centering
\includegraphics[width=.5\textwidth]{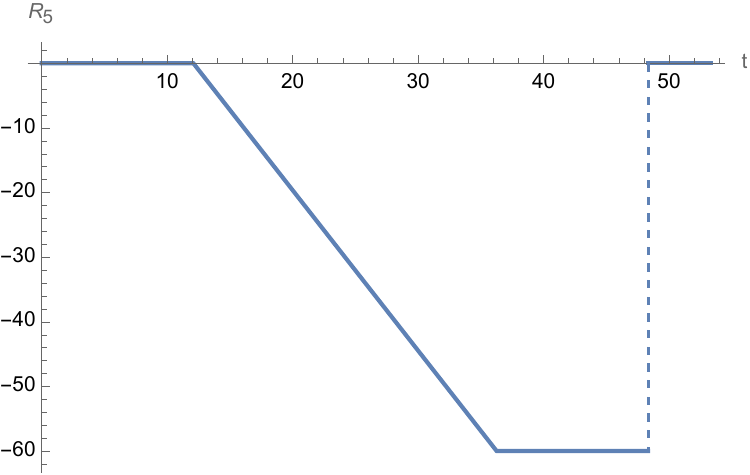}
\qquad
\caption{Time evolution of $R_5(A:B:C:D)$~\eqref{R5EABCDE}. Here, we plot for $\ell=15$. \label{fig:R5EABCDE}}
\end{figure}

This gives qualitatively similar behaviour to  $R_3(A:B)$ in Section \ref{sec:r3twor}. Notice that $R_5\leq 0$. In the vacuum AdS$_3$ case studied in~\cite{Balasubramanian:2024ysu}, $R_5$ is mostly negative, but with regions of the moduli where $R_5$ can be positive. It would be interesting to investigate the sign patterns of $R_5$ for more generic subsystem sizes here. 

We can generalise the $R_5$ calculation to $R_n$ for arbitrary odd $n$ ($n\geq 3$). Likewise, for arbitrary even $n$ ($n\geq 4$), we can generalise the $I_3=\frac{1}{2}I_4$ calculation in Section~\ref{sec:I3} to compute the $n$-partite entanglement signal $I_{n-1}=\frac{1}{2}I_n$. As $I_n$ are functions of von Neumann entropy, their time evolution will be continuous. 


\section{Generalised entanglement membrane for CFT$_2$}
\label{app:gen} 
So far, we have been studying multiparty entanglement signals in the membrane effective theory reviewed in Sect.~\ref{sec:Membrane}. For the special case of 2d CFT, however, that membrane theory becomes degenerate~\cite{Mezei:2018jco}. This is manifested in the bulk by the fact that in the dual BTZ black string~\cite{Hartman:2013qma}, the special extremal slice $z_*$ is pushed to the orbifold singularity, and the two-sided geodesics can penetrate arbitrarily deep into the BTZ interior. This indicates that the bulk $z=e^{\xi}$ coordinate now becomes important and a different scaling ansatz respecting this additional degree of freedom is needed. To account for this, a generalised membrane theory for CFT$_2$ was introduced in~\cite{Jiang:2024tdj}, where the scaling
\es{CoordScalingGen}{
v\rightarrow \Lambda v, \hspace{0.5cm} x \rightarrow \Lambda x, \hspace{0.5cm} \xi\rightarrow \Lambda \xi\,,
}
was adopted. Using this scaling limit in holography, the dynamics of the HRT surface in the bulk BTZ black hole was reduced to an effective membrane theory with \emph{two} degrees of freedom $(x(v),\xi(v))$. While the additional degree of freedom $\xi(v)$ does not change the entanglement entropy value in the scaling limit~\eqref{CoordScalingGen}, it modifies the shape of the membrane $x(v)$. It was conjectured in \cite{Jiang:2024tdj} that in field theory, $\xi(v)$ is related to the infinite-dimensional conformal symmetry.

\begin{figure}[htbp]
\centering
\includegraphics[width=.4\textwidth]{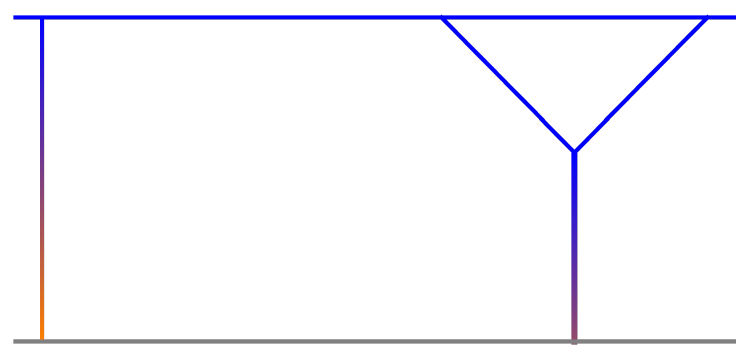}\hspace{0.5cm}
\includegraphics[width=.4\textwidth]{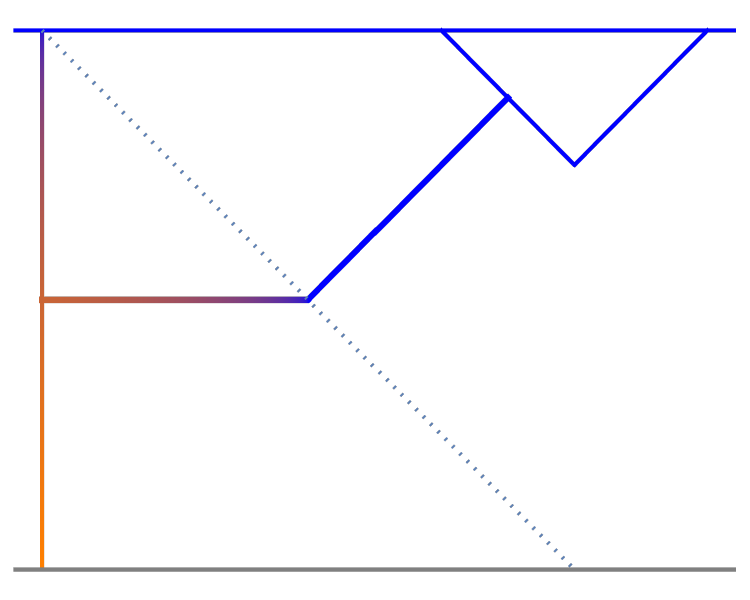}
\qquad
\caption{Membranes computing $S_R$ (thick) at early (left) and late (right) time in generalised membrane theory for CFT$_2$. The dependence on the $\xi$ coordinate is indicated by the colour of the curve, interpolating between blue and orange as one moves from $\xi=O(1)$ to the deep interior. \label{fig:GenMembraneEWCS}}
\end{figure}

The additional degree of freedom $\xi(v)$ has more profound effects on the dynamics of reflected entropy. As in the higher-dimensional cases, there are different candidates for the entanglement wedge cross-section, as pictured in Figure \ref{fig:GenMembraneEWCS}. The vertical one again produces a linear growth in the reflected entropy. In the other candidate, the diagonal segment gives an extensive contribution, but this now only extends across half of the region. The horizontal component does not contribute to the reflected entropy in the scaling limit. As a result, in the generalised membrane theory we have 
\begin{equation}
\label{GenMembraneSR}
\frac{S_R(A:B)}{2} = s_{\text{th}}
\begin{cases}
0 \ &t < \frac{\ell_C}{2}\,, \\ 
t - \frac{\ell_C}{2}\ \hspace{0.5cm} &\frac{\ell_C}{2} < t < \frac{\ell_A+\ell_C}{2}\,, \\ 
\frac{\ell_A}{2}  \  &t > \frac{\ell_A+\ell_C}{2}\,.
\end{cases}
\end{equation}
The height of the plateau is half as big as in the ordinary membrane theory. Note also that we always have $v_E=v_B$ in $d=2$, so $S_R(A:B)$ is now continuous at $t=\frac{\ell_C}{2}$. 

\subsection{$S_R$ in the case of adjacent intervals}
\begin{figure}[htbp]
\centering
\includegraphics[width=.4\textwidth]{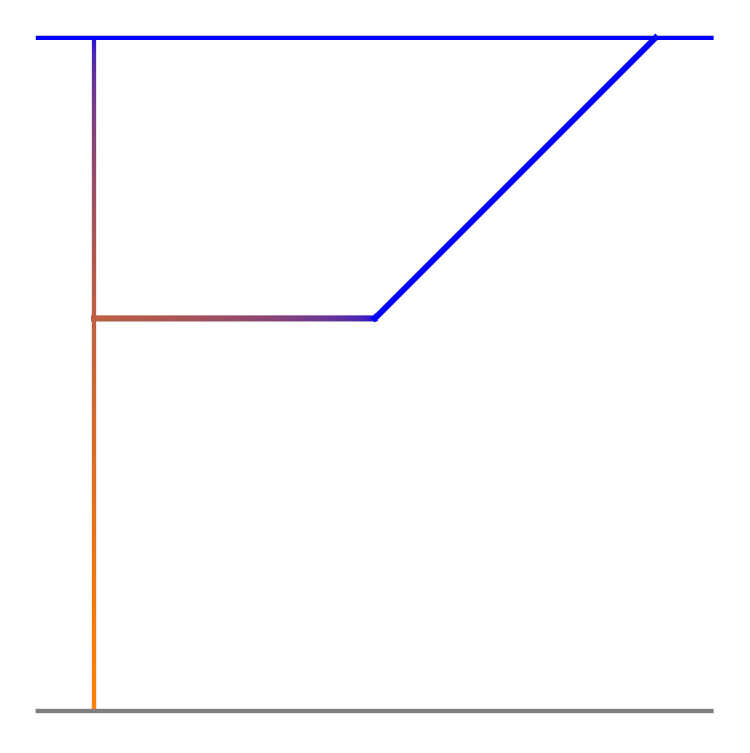}\hspace{0.5cm}
\qquad
\caption{The generalised membranes computing saturated $S_R$ (thick) in 2d CFT in the case of adjacent intervals. As in ~\cite{Jiang:2024tdj}, we use colour to denote the additional degree of freedom $\xi$ on the membrane: the change from blue to orange denotes the increase in $\xi$.\label{fig:GenMembraneSRAdj}}
\end{figure}

The generalised membrane theory for reflected entropy~\cite{Jiang:2024tdj} was developed for the case depicted in Figure~\ref{fig:GenMembraneEWCS}, where an interval of $O(\Lambda)$ size was inserted between $A$ and $B$. To make our multipartite entanglement signal calculations comparable with the ordinary membrane results in Sections~\ref{sec:r3} and~\ref{sec:I3}, we need to develop an effective description for adjacent $A$ and $B$. We assume that as in the disjoint interval case studied in~\cite{Jiang:2024tdj} and reviewed above, the generalised membrane contains a horizontal component that does not contribute to the reflected entropy in the scaling limit. In~\cite{Jiang:2024tdj}, the equations of motion for these generalised membranes parametrized by $(v(x),\xi(x))$\footnote{A change of variable from $v$ to $x$ has been made to describe the horizontal portion of the generalised membrane.} in the scaling limit were found to be
\es{xiEOMx}{
 \xi''&=1-\xi'^2 - e^{4(\xi_p-\xi)}\,,\\
 v'&=e^{-2(\xi-\xi_p)}\,.
}
where $f'=\frac{df}{dx}$. The equations~\eqref{xiEOMx} have the following approximate solutions in the scaling limit
\begin{equation}
    \label{XiEWCS}
    (v(x),\xi(x))=
    \left\{
    \begin{aligned}
        &(v_p,-x+x_0+\xi_p),&& 0<x<x_0\\
        &(x+v_p-x_0,\xi_p),&& x_0<x
    \end{aligned}
    \right.
\end{equation}
where $v_p$, $\xi_p$, and $x_0$ are constants, with $x_0$ labeling the location of the membrane's turning point. Like in ordinary membrane theory~\cite{Jonay:2018yei,Mezei:2018jco}, $v(x)$ describes the shape of the generalised membrane. To visualise the additional degree of freedom $\xi(x)$, we label it as colour on the generalised membrane as in~\cite{Jiang:2024tdj}, see Figure~\ref{fig:GenMembraneEWCS} and~\ref{fig:GenMembraneSRAdj}. There are two boundary conditions for the EWCS: that $x=\ell_A$ at time $t$, and that it ends on the vertical membrane at $x=0$ perpendicularly. Notice that in imposing the boundary conditions for the EWCS in generalised membrane theory, we need to match not only the shape of the membrane $v$, but also the colour on the membrane $\xi$. Written explicitly, the two boundary conditions are 
\begin{align}
    &x_0+\xi_p=t-v_p\,&&\text{to match the vertical membrane HRT,}\label{matchingConds1}\\
    &(\ell_A+v_p-x_0, \xi_p)=(t, 0)\,&&\text{to match the connection point on the $t$ slice}\label{matchingConds2}
\end{align}
where the vertical membrane in~\eqref{matchingConds1} is described by~\cite{Jiang:2024tdj}
\begin{align}
    (x(v),\xi(v))=(0,t-v)
\end{align}
One can then solve for the 3 constants
\begin{align}
    \xi_p=0 && v_p=t-\frac{\ell_A}{2} && x_0=\frac{\ell_A}{2}
\end{align}
The length of the EWCS is thus 
\es{ECWS}{
\ell(\text{EWCS})=x_A-x_0={\ell\ov 2}\,.
}
See Figure~\ref{fig:GenMembraneSRAdj} for the plot of this generalised membrane. Notice that the saturation value is half as big compared to that in ordinary membrane theory as shown in Figure~\ref{fig:SRABC} right. 

As such, the time evolution of $S_R(A:B)$ is given by\footnote{As in the $d>2$ case, when considering adjacent intervals $S_R$ is not UV finite~\cite{Kudler-Flam:2020url}, and we subtract the logarithmic UV divergences~\eqref{AreaLogLaw}. }
\begin{equation}
\label{GenMembraneSRAdj}
\frac{S_R(A:B)}{2} = s_{\text{th}}
\begin{cases}
t \ &t < \frac{\ell_A}{2}\,, \\ 
\frac{\ell_A}{2}  \  &t > \frac{\ell_A}{2}\,.
\end{cases}
\end{equation}

As we will see shortly, the difference in the saturation of $S_R$  between the ordinary and generalised membrane theories,~\eqref{SRABSingReg} vs~\eqref{GenMembraneSRAdj}, leads to strikingly different behaviours of multipartite entanglement signals in 2d CFT and generic chaotic systems. This indicates that the infinite-dimensional conformal symmetry might have substantial effects on multiparty entanglement, a possibility that would be very interesting to investigate in the future.


\subsection{Dynamics of $R_3$}
We use the generalised membrane theory for reflected entropy~\eqref{GenMembraneSRAdj} to study the dynamics of residual information in the setups considered in Section~\ref{sec:r3}. Quantities based solely on von Neumann entropy can be obtained from those in ordinary membrane theory in section~\ref{SingleFiniteReg} by simply setting $v_E=v_B=1$.\footnote{We note that while setting $v_E=v_B=1$ in ordinary membrane theory gives the correct generalised membrane theory results, the membranes computing entanglement entropy in generalised membrane theory have 2 degrees of freedom~\cite{Jiang:2024tdj}. However, the extra degree of freedom will not play any role in determining the value of entanglement entropy if we are not studying the saturation of $S_R$.} 

\subsubsection{A single finite interval}
We consider the setup of a single finite interval, $A = (-\infty, 0)$,  $B = (0, \ell)$, and  $C= (\ell, \infty)$, as is studied in section~\ref{SingleFiniteReg}. $I(A:B)$ and $I(A:C)$ can be simply obtained from~\eqref{IABSingleFiniteReg} and~\eqref{IACSingleFiniteReg} by setting $v_E=v_B=1$. Taking the difference between $S_R(A:B)$~\eqref{GenMembraneSRAdj} and $I(A:B)$, we find that
\begin{align}
    R_3(A:B)=0\ \ {\rm for}\ \ t\in[0,+\infty)
\end{align}
which is different from the ordinary membrane result~\eqref{R3ABSingInt} when setting $v_E=v_B=1$ due to the difference in saturation of $S_R$ in generalised membrane theory. 

$S_R(A:C)$ is computed by the same vertical membrane (with an additional degree of freedom) as is shown in Figure~\ref{fig:SRABCac}, which never saturates. As such, $S_R(A:C)$ is given by~\eqref{SRACSingleFiniteReg} with $v_E=v_B=1$. Taking the difference with $I(A:C)$, we have 
\begin{align}
    R_3(A:C)=0\ \ {\rm for}\ \ t\in[0,+\infty)
\end{align}
Notice that we have $R_3(A:C)=0$ when setting $v_E=v_B=1$ in~\eqref{R3ACSingInt}.  

As $R_3$ is only a signal of tripartite entanglement instead of a measure, $R_3(A:B)=R_3(A:C)=0$ in this setup does not necessarily indicate that there is only bipartite entanglement at any time. 

\subsubsection{Two finite regions}
We consider the setup of two finite regions, $A = (0,\ell_A)$, $B= (\ell_A, \ell_A + \ell_B)$, and $C = (-\infty,0) \cup (\ell_A + \ell_B,\infty)$, as  studied in Section~\ref{sec:r3twor}. Again, $I(A:B)$ and $I(A:C)$ can be simply obtained from~\eqref{IABTwoFiniteReg} and~\eqref{IACTwoFiniteReg} by setting $v_E=v_B=1$. Since we focus on the $\ell_B < \ell_A$ case, using~\eqref{GenMembraneSRAdj}, $S_R(A:B)$ is given by
\begin{equation}
\frac{S_R(A:B)}{2} = s_{\text{th}}
\begin{cases}
t \ &t < \frac{\ell_B}{2}\,, \\ 
\frac{\ell_B}{2}  \  &\frac{\ell_B}{2} < t <\frac{\ell_A+\ell_B}{2} \,.\\
0  \  &t > \frac{\ell_A+\ell_B}{2}\,.
\end{cases}
\end{equation}
Notice that when $AB$ saturates, $S_R(A:B)$ abruptly drops to 0. The corresponding cross-section is shown in Figure~\ref{fig:SatEWCSAdj}. Just as in the ordinary membrane theory case~\eqref{2srab}, this cross-section is no longer described by generalised membrane theory anymore, as it does not grow with $t$ or $x$ in the scaling limit~\eqref{CoordScalingGen}. See Appendix~\ref{SatSRAdj} for details. 
A related calculation on the dynamics of $S_R$ with two finite regions was studied in~\cite{Kudler-Flam:2020url} from both 2d CFT and holographic points of view.

Taking the difference with $I(A:B)$ we find that 
\begin{equation}
R_3(A:B) = s_{\text{th}}
\begin{cases}
0 \ &t < \frac{\ell_A}{2}\,, \\ 
2t-\ell_A  \  &\frac{\ell_A}{2} < t <\frac{\ell_A+\ell_B}{2} \,.\\
0  \  &t > \frac{\ell_A+\ell_B}{2}\,.
\end{cases}
\end{equation}
Notice that ${\rm max}\{R_3(A:B)\}=\ell_B$, which is $\emph{half}$ the bound for $R_3(A:B)$~\cite{Balasubramanian:2024ysu}. As a comparison, in the same setup in ordinary membrane theory, $R_3(A:B)$~\eqref{R3ABTwoReg1} does saturate the $2\ell_B$ bound. 

$S_R(A:C)$ is computed by the same vertical membrane as is shown in Figure~\ref{fig:SRCABCac} (again, with an additional degree of freedom). Hence, we can set $v_E=v_B=1$ in~\eqref{2srac} to obtain $S_R(A:C)$ in generalised membrane theory. Taking the difference with $I(A:C)$, we arrive at 
\begin{equation}
\label{R3ACTwoReg2d}
R_3(A:C) = s_{\text{th}} 
\begin{cases}
0 &  t < \frac{\ell_A}{2} \,, \\
2  t - \ell_A  & \frac{\ell_A}{2} <  t < \frac{\ell_A}{2} +\frac{\ell_B}{4} \,,  \\
\ell_A + \ell_B - 2 t & \frac{\ell_A}{2} +\frac{\ell_B}{4}<  t < \frac{\ell_A+\ell_B}{2} \,,  \\
0 &   t > \frac{\ell_A+ \ell_B}{2} \,.
\end{cases}
\end{equation}
which is~\eqref{R3ACTwoReg} with $v_E=v_B=1$. We find that the initial plateau due to the saturation of $B$ in~\eqref{R3ACTwoReg} disappears in~\eqref{R3ACTwoReg2d}. The maximal value of $R_3(A:C)$ is ${\rm max}\{R_3(A:C)\}=\frac{\ell_B}{2}$, which reaches half the $R_3(A:C)$ bound. 

\subsection{Dynamics of $Q_4$}
We use the generalised membrane theory for reflected entropy~\eqref{GenMembraneSRAdj} to study the dynamics of $Q_4$ in the setup considered in Section~\ref{sec:I3}: $A = (0, \ell_A)$, $B = (\ell_A, \ell_A+\ell_B)$, $C= (\ell_A+\ell_B, \ell_A + \ell_B+\ell_C)$ and $D = (-\infty,0) \cup (\ell_A + \ell_B + \ell_C, \infty)$. Again, we assume $\ell_A=\ell_C$ for simplicity. 

The $\ell_A<\ell_B$ case is similar to that considered in section~\ref{sec:I3} except for the difference in saturation: the saturated EWCS ends on the vertical membrane on the left, but the saturation value is half as big as that in~\eqref{Q4DABCD1}. Therefore, we have  
\begin{equation}
\label{Q4DABCD2d1}
Q_4 = s_{\text{th}} 
\begin{cases}
0 \ &0<t < \frac{\ell_A+\ell_B}{2}\,, \\ 
2 \ell_A &\frac{\ell_A+\ell_B}{2}<t < \frac{\ell_A+\ell_B+\ell_C}{2}\,, \\ 
0\ &t>\frac{\ell_A+\ell_B+\ell_C}{2}\,.
\end{cases}
\end{equation}
Compared to the corresponding ordinary membrane theory result~\eqref{Q4DABCD1}, we see that ${\rm max}\{Q_4\}=2\ell_A$ reaches half the $2\ell_A$ bound.  

\begin{figure}[htbp]
\centering
\includegraphics[width=.31\textwidth]{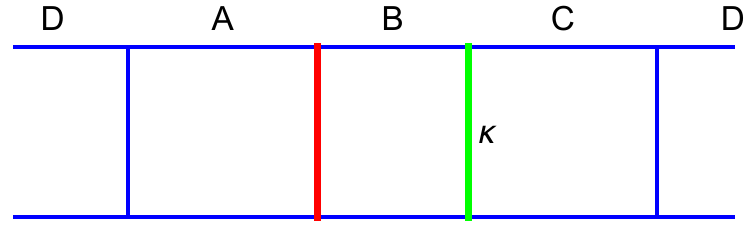}\hspace{3mm}
\includegraphics[width=.31\textwidth]{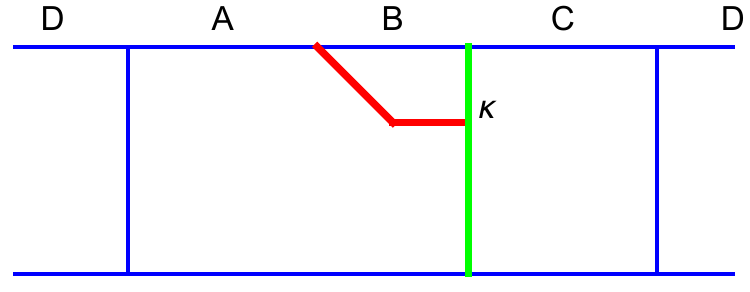}\hspace{3mm}
\includegraphics[width=.31\textwidth]{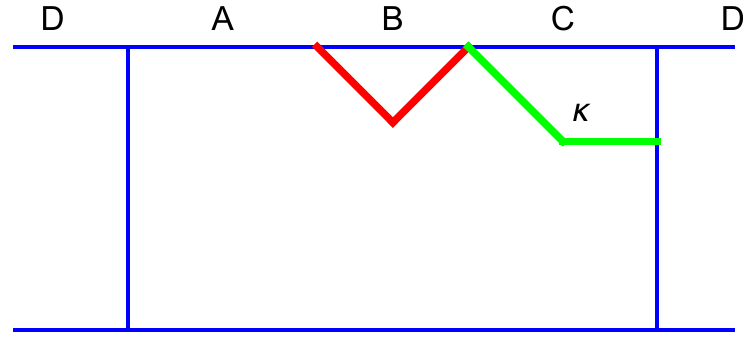}
\qquad
\caption{Minimal cross-sections computing $S_{AA^*A_*A^*_*}$ (red, thick) in $Q_4$ in the generalised membrane theory for CFT$_2$ when $\ell_A>\ell_B$. Notice that the saturation of $\kappa$ (green, thick) changes the shape as well as the value of the membrane computing $S_{AA^*A_*A^*_*}$. Here, we do not plot the additional degree of freedom $\xi$ as a colour on the membrane as in Figure~\ref{fig:GenMembraneSRAdj}, instead we follow the same colour conventions for $Q_4$ in the ordinary membrane cases in e.g. Figure~\ref{fig:kappa}. \label{fig:Q4DABCDLgA2d}}
\end{figure}

The $\ell_A>\ell_B$ case is more interesting: since the saturated EWCS now ends on $\kappa$, and the saturation value is half as big compared to the ordinary membrane result, the saturation of $\kappa$ does change the value of $S_{AA^*A_*A^*_*}$ by doubling its value, see Figure~\ref{fig:Q4DABCDLgA2d}. This is in contrast to the ordinary membrane theory case shown in Figure~\ref{fig:kappa}, where the saturation of $\kappa$ does not change the value of $S_{AA^*A_*A^*_*}$.

We thus have
\begin{equation}
\label{Q4DABCD2d2}
Q_4 = s_{\text{th}} 
\begin{cases}
0 \ &0<t < \frac{\ell_A+\ell_B}{2}\,, \\ 
2 \ell_B &\frac{\ell_A+\ell_B}{2}<t < \ell_A\,, \\ 
4 \ell_B &\ell_A<t < \frac{\ell_A+\ell_B+\ell_C}{2}\,, \\ 
0\ &t>\frac{\ell_A+\ell_B+\ell_C}{2}\,.
\end{cases}
\end{equation}
Notice that unlike the $\ell_A<\ell_B$ case, now ${\rm max}\{Q_4\}=4\ell_B$ does saturate the $4\ell_B$ bound. This indicates that the extensive part of the entanglement of $B$ is entirely multipartite. 

\section{Discussion}
\label{sec:discussion}

We have considered  time evolution of multiparty entanglement signals in the late time, large region limit, where  time dependence can be calculated using the effective membrane theory introduced in \cite{Nahum:2016muy,Jonay:2018yei,Mezei:2018jco}. This limit describes the behaviour of a broad class of chaotic quantum many-body systems, including some that have holographic gravity duals. Our calculations that just involve von Neumann entropies, like the triple information $I_3$, apply to all these systems.  Some of our entanglement signals also involve the reflected entropy, and we used the membrane picture of  \cite{Jiang:2024tdj} to study this quantity.


Our results show that multiparty entanglement signals have several striking dynamical behaviours.
First, different signals have different kinds of dynamics. This finding strengthens the arguments of \cite{Balasubramanian:2024ysu} that no one quantity is going to capture all aspects of multiparty entanglement. This sort of entanglement structure is more complicated than bipartite entanglement (see, e.g., \cite{Verstraete:2002gqj}), and we are likely to need multiple different scalar quantities to characterise different types of multiparty entanglement. In particular, the fact that there are periods where one signal is constant while others are changing (see, for example, Figure~\ref{fig:I3Q41}) suggests that they measure different things, and are not simply different functions of a common feature of the state.

Second, the signals that involve reflected entropy can change discontinuously in time. In fact, it is known that reflected entropy can change discontinuously in holographic states as the separation between entangled regions varies.  This happens because the size of the associated entanglement wedge cross-section can change discontinuously  \cite{Dutta:2019gen}. Here, we are finding that discontinuities can appear dynamically as a consequence of physical processes in the hydrodynamic limit. This suggests dynamical phase transitions in the entanglement structure. The discontinuity in the signals arises from a change in the dominant saddle for the canonical purification, which gives a discontinuity in the reflected entropy. This is similar to the Hawking-Page transition~\cite{Hawking:1982dh}, where there is a discontinuity in the entropy which similarly arises from an exchange of saddles. In the Hawking-Page transition, we can identify the phase transition in the dual as a confinement-deconfinement phase transition. In the present case we do not have a good understanding of the dual description. But as for the Hawking-Page transition, we have a discontinuity only in the infinite $N$ limit; at finite $N$ both saddles contribute and the transition is smoothed out. This thermodynamic infinite $N$ nature of the discontinuity suggests that it is reflecting a phase transition in the dual theory, and not just some feature of the definition of the signals. It would be interesting to understand the physics of these  transitions.\footnote{Similar discontinuities in the dynamical evolution of subregion complexity were recently observed in \cite{Fan:2025moc,Haah:2025hyf}.}


In the limit we consider, the multiparty signals are always initially zero. This reflects the fact that the state  initially has entanglement only on scales small compared to the regions we consider, and it takes some time for the entanglement to spread to larger scales and potentially contribute to our multiparty signals. More surprising was that in many cases the signals are also zero in the late time equilibrium. We might have expected that multiparty entanglement would increase monotonically as the system evolved to more generic states. However, this expectation is too naive. If we consider, for example, a random state on three parties $A$, $B$, $C$, when $d_C > d_A d_B$ (where $d_X = \mathrm{dim} (\mathcal H_X)$ is the dimension of the corresponding Hilbert space), the reduced density matrix $\rho_{AB}$ is approximately maximally mixed, and hence factorised: $\rho_{AB} \approx \frac{1}{d_A d_B} \mathbb{I}_A \otimes \mathbb{I}_B$. This indicates that the entanglement is bipartite between $C$ and $A$, and bipartite between $C$ and $B$. If the evolution in the time-dependent system drives the system to more generic states, one might expect the entanglement structure in such cases to approach this kind of bipartite structure at late times, which is consistent with what we find. 

We saw that in some cases, the state saturates bounds on  entanglement signals established in \cite{Balasubramanian:2024ysu}, indicating that in one of the regions the entanglement structure is entirely  multipartite. This provides further evidence for the general importance of multiparty entanglement; in the simple and universal dynamical context that we consider, the amount of multiparty entanglement can be saturated.

Finally, we observed that the time evolution profiles of multiparty entanglement signals are different in 2d CFT. This difference can be traced back to the fact that the effective theory underlying entanglement dynamics in 2d CFT is the generalised membrane theory, which contains an additional degree of freedom compared to ordinary membrane theory and hence posits a different dynamical behaviour of reflected entropy~\cite{Jiang:2024tdj}. These differences indicate that the infinite-dimensional conformal symmetry in two dimensions might play a key role in the dynamics of multiparty entanglement. The generalised membrane theory~\cite{Jiang:2024tdj} was derived using holography. It would be interesting to understand this effective theory from a field-theoretic point of view. 

Our  membrane calculations of multipartite entanglement signals  are appropriate to chaotic systems. It would be instructive to compare these results with the same quantities in integrable models, whose entanglement dynamics are described by the quasi-particle picture~\cite{Calabrese:2005in,Casini:2015zua}. Assuming no initial multipartite entanglement,\footnote{See~\cite{Casini:2015zua} for discussions on the quasiparticle picture with initial multiparty entanglement.} the entanglement dynamics in the quasiparticle picture are bipartite by construction, and involve free-streaming EPR pairs of entangled quasiparticles which can be shared by at most two regions at a time.  One would therefore expect multipartite entanglement signals to vanish for integrable systems. Indeed, in rational conformal field theory (RCFT), it has been shown that $R_3=0$ in 3-party states~\cite{Kudler-Flam:2020url} and  $I_3=0$ in 4-party states~\cite{Balasubramanian:2011at} at all times. The latter result follows directly from the formula~\cite{Calabrese:2005in} 
\begin{align}
\label{EERCFT}
    S_A(t)\sim S_A(\infty)+\frac{\pi c}{12 \epsilon}\sum_{k,\ell=1}^{2 n}(-1)^{k-\ell-1}\text{max}(u_k-t,u_\ell+t)
\end{align}
for $n$ intervals. Likewise, the six-party entanglement signal $I_5$~\eqref{ninf} also vanishes in a 6-party state: Suppose we have a system $|\psi\rangle_{\prod_{j=1}^6A_j}$ with subregions $A_j=(\sum_{k=0}^{j-1}\ell_k,\sum_{k=0}^j\ell_k)$ ($1\leq j\leq 5$) and $A_6=(-\infty,0)\cup (\sum_{k=1}^5\ell_k,+\infty)$, and perform a trace over $A_6$. Using~\eqref{ninf} for RCFTs, where each combination $S_i$ consists of $i$-party entropy computed from~\eqref{EERCFT}, it is straightforward to show that 
\begin{align}
    I_5(A_1:\ldots:A_5)=0\ \ {\rm in}\ \ |\psi\rangle_{\prod_{j=1}^6A_j}
\end{align}
Since, in the absence of initial (bipartite or multipartite) entanglement,  the quasiparticle picture leads to purely bipartite entanglement while, as we showed,  the membrane picture often produces non-trivial multipartite entanglement, we can regard the generation of multiparty entanglement as a common feature of scrambling dynamics. It would be interesting to investigate cases with initial multipartite entanglement in the quasiparticle~\cite{Casini:2015zua,Berthiere:2024sio} and membrane pictures in future works. 

The membrane theory of reflected entropy that we used was derived holographically by examining entanglement wedge cross-sections. We expect the membrane picture to apply more generally to chaotic systems, but this has not yet been shown. The authors of ~\cite{Akers:2021pvd,Akers:2022zxr,Akers:2024pgq} studied reflected entropy in random tensor networks in a time-symmetric setting, and showed that it was computed by the analog of an entanglement wedge cross-section.  If we could find a way to define sensible time-dependent tensor networks, we could perhaps extend the approach of \cite{Akers:2021pvd,Akers:2022zxr,Akers:2024pgq} to understand the dynamics of multiparty entanglement in states constructed by tensor networks, and perhaps to test whether the membrane picture of entanglement that we used applies more generally.

A comparison between the residual information $R_3(A:B)$ and the genuine multi-entropy $GM(A:B:C)$ in vacuum AdS$_3$, showed that time-symmetric holographic states cannot have purely GHZ-like entanglement \cite{Balasubramanian:2025hxg}.  It would be interesting to explore whether this statement remains true in time-dependent holographic setups or chaotic states more generally. To proceed, one needs first to understand the time evolution of $GM(A:B:C)$ in the scaling limit; the techniques we developed in this paper should be helpful towards this goal.

Finally, it would be interesting to understand whether the spread of multiparty entanglement  contributes in some way to the growing complexity of the quantum state over time.  It would be reasonable  to suppose that this could be the case, because constructing specific entanglement patterns between multiple parties using just local operations requires a complicated quantum circuit \cite{Balasubramanian:2018hsu}.  On the other hand, the von Neumann entropy in thermalizing states saturates well before the complexity does \cite{Susskind:2014moa}.   But perhaps sufficiently complex patterns of multi-party entanglement do not saturate in this way, especially if the number of parties grows exponentially large in the entropy of the gravitating system, likely engaging non-perturbative corrections to the standard geometrisation of entanglement entropy.

\appendix

\section{Saturation of $S_R$ in finite adjacent intervals}\label{SatSRAdj}
In this Appendix, we use holography to verify that the saturated $S_R$ in the case of finite adjacent intervals as shown in Figure~\ref{fig:SatEWCSAdj} in the main text is indeed a vertical ``membrane'' that has vanishing extensive entropy in the scaling limit. 

We calculate saturated $S_R$ in holographic CFTs by examining the entanglement wedge cross-sections in the dual gravitational theories, where the planar AdS$_{d+1}$-Schwarzschild black brane metric is given by
\begin{align}
    ds^2=\frac{1}{z^2}\Big(-f(z)dt^2+\frac{dz^2}{f(z)}+dx^2+ d\vec{y}_{d-2}^2\Big), \ \ \ f(z)=1-z^d\label{AdSSchMetric}
\end{align}
The Schwarzschild time $t$ in~\eqref{AdSSchMetric} is related to the infalling time $v$ in~\eqref{blackBraneInf} via
\begin{align}
\label{SchInf}
    v(z)=t-\int_0^z\frac{dz'}{f(z')}=t-\frac{1}{d}B\left(z^d;\frac{1}{d},0\right),\ \ z\in [0,1)
\end{align}
where $B\left(z;a,b\right)$ is the incomplete Euler Beta function. The saturated EWCS lies on a constant Schwarzschild time slice. 

We are interested in the bulk RT surface homologous to a boundary subregion $AB$ of size $X=\ell_A+\ell_B$, see Figure~\ref{fig:SatEWCSAdj} (top). Like in section~\ref{sec:r3twor}, we consider the case $\ell_A>\ell_B$, with the boundary point $K$ between $A$ and $B$ located at $x=k>0$. We scale $\ell_A, \ell_B$ and $k$ to be large.

For simplicity, let us first consider the $d=2$ special case.\footnote{Notice that for saturated $S_R$, all the extremal surfaces involved are outside the black hole horizon. Hence, in this case the additional degree of freedom $\xi$ in generalised membrane theory is always of $O(1)$ value and can be scaled away. The results are thus the $v_E=v_B=1$ special case of planar AdS$_{d+1}$-Schwarzschild black holes.} In~\cite{Jiang:2024tdj}, it was shown that the static spacelike geodesic homologous to a finite boundary subregion is given by the parametric curve $\big(z(\lambda),x(\lambda)\big)$
\begin{align}
\begin{split}
\label{statGeod}
    z(\lambda)&=\frac{\sinh \frac{X}{2}}{\sqrt{\cosh \lambda \cosh (X+\lambda )}}\\
    x(\lambda)&=\frac{1}{2} \log (\text{sech}\lambda  \cosh (X+\lambda ))
\end{split}
\end{align}
where $\lambda\in(-\infty,+\infty)$. To find the minimal cross-section bisecting the $AB$ entanglement wedge, we minimise the geodesic distance between $K$ \footnote{To regularise the UV divergence, we consider the point $K$ at $z=\epsilon\ll 1$ in the asymptotic region.} and an arbitrary point $G$ on the parametric curve~\eqref{statGeod}. The geodesic distance formula in planar BTZ black brane is~\cite{Mezei:2016zxg,Shenker:2013pqa} 
\begin{align}
    \cosh d(K,G)=T_1^KT_1^G+T_2^KT_2^G-X_1^KX_1^G-X_2^KX_2^G\label{geodDistBTZ}
\end{align}
where the embedding coordinates are given by
\begin{align}
\begin{split}
    T_1&=\frac{\sqrt{1-z^2}}{z} \sinh t,\ \ \ \ T_2=\frac{1}{z}\cosh t\\
    X_1&=\frac{\sqrt{1-z^2}}{z} \cosh t,\ \ \ X_2=\frac{1}{z}\sinh x
\end{split}
\end{align}
The explicit expression for the geodesic distance between $K$ and $G$ is long and uninspiring. In the large region limit, we find that the minimal geodesic distance, after dropping the UV divergent $\log\frac{2}{\epsilon}$ term, is 
\begin{align}
    d_{{\rm reg}}=0 \ \ \ {\rm at} \ \ \ \lambda_*=-\frac{X}{2} + k
\end{align}
At $\lambda_*$, one finds 
\begin{align}
    x(\lambda_*)=k
\end{align}
In other words, the minimal geodesic between $K$ and $G$ is the $\emph{radial}$ one in the scaling limit. This result makes sense intuitively, as the geodesic~\eqref{statGeod} contains a plateau region where it skims the horizon almost transversely along the $x$ direction, and the EWCS needs to end on it perpendicularly. 

Now, let us consider saturated $S_R$ in the adjacent interval case in $d>2$ with strip entangling subregions. In the large region limit, the intersection point $z_0$ between the EWCS and the RT surface is close to the horizon, $z_0=1-\delta$ ($\delta\ll 1$). Following the intuition  above, the EWCS is also almost radial in the scaling limit when $d>2$.  Thus, it would be convenient to parametrise the RT surface by $x(z)$ (instead of $z(x)$). The area functional is given by
\begin{align}
    A=\int dz d^{d-2}\vec{y}\ \frac{1}{z^{d-1}}\sqrt{\frac{1}{f(z)}+x'(z)^2}
\end{align}
Noting that $x(z)=k$ and $x'(z)=0$ for the saturated EWCS, the area functional becomes
\begin{align}
\label{AreaSatEWCSdGeq2}
    A&=\int_{\epsilon}^{z_0} dz  \frac{1}{z^{d-1}}\frac{1}{\sqrt{1-z^d}}=\frac{1}{d}B\left(z^d;\frac{2}{d}-1,\frac{1}{2}\right)\Big|_{\epsilon}^{z_0}\\
    &=\frac{\sqrt{\pi } \Gamma \left(\frac{2}{d}-1\right)}{d \Gamma \left(\frac{2}{d}-\frac{1}{2}\right)}-\frac{2}{\sqrt{d}}\sqrt{\delta}+\frac{1}{d-2}\frac{1}{\epsilon ^{d-2}}
\end{align}
where we have eliminated the $\int d^{d-2}\vec{y}={\rm Area}(\p A)$ infinite factor as in the main text after~\eqref{AreaFunct2}. Restoring this factor, we find that the last term diverges as $\frac{1}{d-2}\frac{{\rm Area}(\p A)}{\epsilon ^{d-2}}$, which is the same as the UV divergence of RT surfaces~\cite{Rangamani:2016dms}.\footnote{This result makes sense, as both the RT surface and the saturated EWCS end on the boundary perpendicularly, and are hence asymptotically the same.} Dropping this UV divergent $\frac{{\rm Area}(\p A)}{\epsilon ^{d-2}}$ term, we have  
\begin{align}
    A_{{\rm reg}}\sim \#\sqrt{\delta}+O(1)\to 0
\end{align}
in the scaling limit. Notice that when $d=2$, the area/length in~\eqref{AreaSatEWCSdGeq2} instead integrates to 
\begin{align}
    A=-\tanh ^{-1}\left(\sqrt{1-z^2}\right)\Big|_{\epsilon}^{z_0}=-\sqrt{2} \sqrt{\delta}+\log\frac{2}{\epsilon}
\end{align}
which agrees with our $d=2$ analysis above.

To sum up, we find that in $d\geq 2$ in the large region limit, the regularised area of the saturated EWCS is $0$. As these EWCS as shown in Figure~\ref{fig:SatEWCSAdj} (top) are radial and reach close to the horizon, their projections to the boundary along constant infalling time are vertical lines dangling from $K$, see Figure~\ref{fig:SatEWCSAdj} (bottom). This is due to the near-horizon logarithmic divergence of infalling time $v$~\eqref{SchInf}~\cite{Jiang:2024tdj}. Nevertheless, these cross-sections are no longer described by membrane theory for holographic CFT~\cite{Mezei:2018jco,Jiang:2024tdj}. This is because to derive the membrane tension~\eqref{MinMemb} from the area of the HRT surface in the scaling limit, one needs to integrate out the radial bulk $z$ degree of freedom. However, from e.g.~\eqref{AreaSatEWCSdGeq2} above, it is clear that for the almost-radial cross-section shown in Figure~\ref{fig:SatEWCSAdj}, the radial bulk $z$ degree of freedom $\emph{cannot}$ be integrated out. The reason why this cross-section fails to be captured by membrane theory in other chaotic systems than holographic CFTs is technically less clear. Intuitively, one can understand the reason as cross-sections of this kind do not grow with $t$ or $x$ in the scaling limit. Hence, these vertical ``membranes'' should not be confused with the vertical membranes with tension $\mathcal{E}(0)=v_E$~\cite{Mezei:2018jco} obtained from projections of the HRT surfaces in the black hole interior~\cite{Hartman:2013qma,Liu:2013iza,Liu:2013qca}. The latter reaches the $t=0$ slice in membrane theory, computing entanglement entropy $S=s_{\rm th} v_E t$ that grows linearly with time.

\acknowledgments
We thank M\'ark Mezei for many helpful conversations.  We are also grateful to Geoffrey Penington, Julio Virrueta, and Gabriel Wong for discussions. HJ is partially supported by Lady Margaret Hall, University of Oxford. VB was supported in part by the DOE through DE-SC0013528 and the QuantISED grant DE-SC0020360, and  by the Eastman Professorship at Balliol College, University of Oxford.  SFR was supported in part by STFC through grant number ST/X000591/1.



\bibliographystyle{JHEP}
\bibliography{biblio.bib}






\end{document}